\documentclass[manuscript, 11pt]{aastex}
\pdfoutput=1
\usepackage{float}
\usepackage{color}
\usepackage{graphicx}

\begin{document}


\title{Optical Spectral Observations of a Flickering White-Light Kernel in a C1 Solar Flare}
\author{Adam~F.~Kowalski}
\affil{NASA Postdoctoral Program Fellow, Goddard Space Flight Center, Code 671, 8800 Greenbelt Rd., Greenbelt, MD  20771 USA}
\email{adam.f.kowalski@nasa.gov\\} 
\and
\author{Gianna~Cauzzi } 
\affil{INAF-Osservatorio Astrofisico di Arcetri, Firenze, Italy \& National Solar Observatory, Sunspot NM, USA}
\and
\author{Lyndsay~Fletcher}
\affil{SUPA School of Physics and Astronomy, University of Glasgow, Glasgow G12 8QQ,  UK}

\begin{abstract}
We analyze optical spectra of a two-ribbon, long duration C1.1 flare that occurred on
18 Aug 2011 within AR 11271 (SOL2011-08-18T15:15). 
The impulsive phase of the flare was observed with a comprehensive set of space-borne and ground-based instruments,
 which provide a range of unique diagnostics of the lower flaring atmosphere.  Here we report
 the detection of enhanced continuum emission,
 observed in low-resolution spectra from 3600 \AA\ to
4550 \AA\ acquired with the Horizontal
Spectrograph at the Dunn Solar Telescope.  A small, $\le$0\arcsec.5 ($10^{15}$ cm$^2$) penumbral/umbral kernel brightens repeatedly in the optical continuum 
and chromospheric emission lines, similar to the temporal characteristics of the hard X-ray variation 
as detected by the Gamma-ray Burst Monitor (GBM) on the Fermi spacecraft.  
Radiative-hydrodynamic flare models that employ a nonthermal electron beam energy flux
high enough to produce the optical contrast in our flare spectra would predict a large Balmer jump in emission, indicative of hydrogen recombination radiation from the upper flare chromosphere. However, we find no evidence of
such a Balmer jump in the bluemost spectral region of the continuum excess.
Just redward of the expected Balmer
jump, we find evidence of a ``blue continuum bump'' in the excess emission which may be indicative of the merging of the higher order Balmer lines.
The large number of observational constraints provides a springboard for modeling the blue/optical emission for this particular flare
with radiative-hydrodynamic codes, which are necessary to understand the opacity effects for the continuum and emission line radiation at these wavelengths.

\end{abstract}


\section{Introduction}

Blue/optical flare spectra contain a wealth of information on the response of the lower stellar atmosphere to flare energy input, which can be diagnosed through line and continuum measurements. However,
spectroscopic observations of the white-light (WL) optical continuum in solar flare kernels are few and far between, dating back to a handful of
spectra obtained over 30 years ago, primarily with the Universal Spectrograph at the Sacramento Peak Observatory \citep[][see also \cite{Hiei1982}]{Machado1974, Neidig1983, Donati1984}.  The few spectra that exist show diverse continuum properties.  \cite{Neidig1983} compiled three of these spectra, looking at the Balmer jump region ($\lambda=3646$ \AA, corresponding to the long-wavelength edge of the recombination continuum onto hydrogen $n=2$). The Balmer jump appeared in one event, not at all in the second and, apparently, 50 \AA\ redward of the Balmer edge wavelength in the third. These spectra have been modeled as a combination of two continuum emission components:  a free-bound hydrogen Balmer recombination spectrum and a H$^-$ recombination spectrum \citep{Hiei1982, Neidig1983, Donati1985, Boyer1985, Neidig1994}. It has also been suggested that optical emission attributed to H$^-$ is entirely due to hydrogen Paschen recombination radiation \citep{Neidig1984}, but \cite{Boyer1985} were unable to find a satisfactory answer from a Paschen (or Balmer) recombination spectrum fit to their flares, and instead argued for H$^-$ emission.  A claimed third continuum component at $\lambda < 4000$ \AA\  \citep{Zirin1980, Zirin1981}, has been shown to result from the merging of Stark-broadened hydrogen line wings, creating pseudo-continuum emission between the Balmer lines,  and between the expected Balmer edge wavelength and the bluest observed Balmer line \citep{Donati1985}.  

Ideas about the lower flare atmosphere obtained from these spectra  have been reached primarily through comparison with static, isothermal, constant density models.  A hydrogen recombination model has been used to infer temperatures of $7000-10,000$ K and electron densities of $\sim 1-5\times10^{13}$ cm$^{-3}$, which gives an origin in the lower chromosphere ($\sim$1000 km above the quiet-Sun $\tau_{5000}=1$ level).  Increased emission from H$^-$ during flares on the other hand implies a temperature increase of the upper photosphere ($\sim$50\,--\,300 km above the quiet-Sun $\tau_{5000}=1$ level) by at least several hundred K \citep[see][for a review]{Neidig1989}. However, the limitation of information derived from these models is that the important line \emph{and} all continuum components are assumed to originate from a common uniform, static, one-dimensional atmospheric layer, a rather crude approximation long discredited by observations \citep[e.g.][]{1996A&A...306..625C,2002A&A...387..678F}.  With this type of analysis, it is not possible to constrain a combination of emission mechanisms from dynamic gradients in the temperature, density, pressure, and ionization.

Among the various heating mechanisms which have been considered to produce the white-light emission, the most commonly-cited is the bombardment of the chromosphere by nonthermal deka-keV electrons \citep{Hudson1972, Abbett1999}, leading either to direct heating of the photosphere or production of free-bound emission in the mid chromosphere \citep[which can also drive UV radiative backwarming, e.g.,][]{Machado, Hawley1992}. Electrons are favoured because of the close relationship between the timing and spatial locations (i.e., kernels) of white-light and hard X-ray emission \citep{Hudson1992}. Energetically, this tends to require all electrons down to around 20~keV to excite the radiation, but it is not clear that these electrons can reach the altitudes required to produce the continuum: certainly not the photosphere, and even reaching the mid chromosphere can be challenging \citep{Fletcher2007}. Other possible heating mechanisms include bombardment by non-thermal MeV protons \citep{Svestka1970, Machado1978}, heating by Alfv\'en waves \citep{Fletcher2008, Russell2013}, or a heated compression wave propagating towards the photosphere \citep{Livshits1981}. However, until the optical spectrum of flares has been properly and systematically characterized, and compared with model predictions \citep[e.g. the radiation hydrodynamics models of ][]{Allred2005} it will not be possible to precisely identify the heating mechanism(s) responsible.

White-light emission was once thought to only originate in large flares, but now has been observed from $\sim$C2 through X-class \citep{Hudson2006, Fletcher2007, Jess2008, Kretzschmar2011}. Unfortunately, the focus on high spatial and temporal resolution in modern solar observations means that almost all current white-light data are solely from narrow-band (e.g., G-band) or broad-band (TRACE/WL) images. There is very little broad-band spectroscopy (color), or information about optical emission line behavior. If the white-light spectrum is known, energetics can be constrained \citep{Neidig1994, KerrFletcher}, and a direct comparison made with the nonthermal particle power deduced from hard X-ray observations \citep{Metcalf2003, Fletcher2007} and with the spectral 
models from each of the proposed heating mechanisms.  In recent times, this has only been done for the Sun using available 3-color (red/green/blue) filter measurements, all at wavelengths longer than the Balmer edge. For example, a recent superposed epoch analysis by \citet{Kretzschmar2011}
of Sun-as-a-Star three-color measurements made at the peaks of flares from upper-C to X-class shows that the data are consistent with a $T\sim$9000 K blackbody continuum.  Using the Hinode Solar Optical Telescope, \citet{KerrFletcher} and \cite{Watanabe2013} found consistency with a much lower temperature blackbody (in the former, free-bound continuum was also possible,  but much more demanding energetically). It is interesting that a hot (9000 K) blackbody continuum component has never been reproduced in radiative-hydrodynamic models that employ a realistic heating model, although it is well-known to dominate the optical spectra and broadband energetics during flares on active M dwarf stars \citep{Hawley1991, Fuhrmeister2008, Kowalski2013}.

In this paper, we present the first spatially and temporally resolved spectra with broad-wavelength  ($\sim$3600\,--\,4550 \AA) and moderate spectral resolution (R $\sim$ 4,000) coverage of a white-light kernel, obtained during a small C-class flare. Simultaneous imaging spectroscopy in photospheric and chromospheric lines allows a clear framing of the white-light emission with respect to the global spatial and temporal development of the flare, which was not possible at the time of the earlier spectroscopic investigations of white-light flares. Indeed, as remarked in \citet{Neidig1989}, older broadband spectra were never obtained on the brightest flaring kernel during the impulsive phase. Further, a modern day investigation of the continuum emission is especially important because of the availability of complementary data in the UV and EUV with the Solar Dynamics Observatory's Atmospheric Imaging Assembly \citep{2012SoPh..275...17L}, as well as nonthermal hard X-rays from the Fermi Gamma-ray Burst Monitor \citep[GBM;][]{Meegan2009}. These data will allow the nonthermal electron energy and number flux to be constrained, to be used as input to future flare models, allowing the heating and excitation mechanisms to be tested.

Section \ref{sec:observation_section} describes the observations and spectral data reduction, and   
Section \ref{sec:WLresults} describes the white-light detection.  In Section \ref{sec:timing}, we present the chromospheric
emission line properties.  In Section \ref{sec:discussion}, we
summarize our observations and discuss some of their physical implications, and how they compare with isothermal slab and hydrodynamic flare models.  Section \ref{sec:conclusions} contains several conclusions from the data.

\section{Observations and Data Reduction} \label{sec:observation_section}

The active region NOAA 11271 (N16.5E42.1) produced a C1.1 flare on August 18, 2011
with a GOES 1\,--\,8 \AA\ peak at approximately UT 15:15 (SOL2011-08-18T15:15).  The flare exhibited one extended ribbon in weak-field plage region, and much more compact, short ribbons or groups of footpoints in the sunspot umbra/penumbra (Figure \ref{fig:fov}). It had a fairly long decay in GOES, with several hard X-ray peaks, but we concentrate here on the largest impulsive burst at around UT 15:09:30.  Unfortunately, RHESSI was in the South Atlantic Anomaly during the main burst and the optical observations, but Fermi registered the event from around 6\,--\,25~keV, allowing for a comparison of the optical data with the X-ray impulsive phase.  

We observed this flare with a comprehensive set of instruments at the
Dunn Solar Telescope (DST) of the National Solar Observatory, employing adaptive optics \citep{2004SPIE.5490...34R}. Region NOAA 11271 was monitored continuously between UT 14:10 and 16:20, with some brief interruptions to re-point the instruments. Atmospheric conditions were clear, and seeing conditions remained good and fairly stable throughout.   The blue light from the DST was directed to the Horizontal Spectrograph
(HSG) with a setup described in Section \ref{sec:hsgsetup}, whereas
the red light was directed to the Interferometric Bidimensional
Spectrometer \citep[IBIS,][]{2006SoPh..236..415C}. Preliminary results have been presented in
\cite{Fletcher2013}, and a comprehensive multi-wavelength analysis
will follow in a subsequent paper.  In this paper, we focus on the WL detection and
the optical emission line characteristics compared to the X-ray impulsive phase.

\subsection{X-Ray Data}
We obtained the Fermi/GBM and GOES 1-8 \AA\ (1.5\,--\,12.4 keV) light curves using the IDL SolarSoft OSPEX software.  
The Fermi/GBM CSPEC file from the NaI detector \#5 (the most sunward facing) was used to produce a  
14.58\,--\,20.70 keV hard X-ray count flux light curve.  This light curve was detrended 
to remove the long-term background modulation, and a small residual pre-flare enhancement was also subtracted. 
 The live time was 4.07~s until 15:08:56, after
which an automatic trigger initiated with a live time of 1.02~s until about 15:19.  We bin all data to 4.07~s
and the count flux is normalized to the peak value of 0.33 counts cm$^{-2}$ s$^{-1}$ keV$^{-1}$ at 15:09:25. The hard X-ray light curve is shown in Figure \ref{fig:fermi_goes} and its properties are described in Sections \ref{sec:WLresults} and \ref{sec:timing}. 

\subsection{IBIS data} \label{sec:IBIS}

\begin{figure}[h]
\centering
\includegraphics[width = 5in]{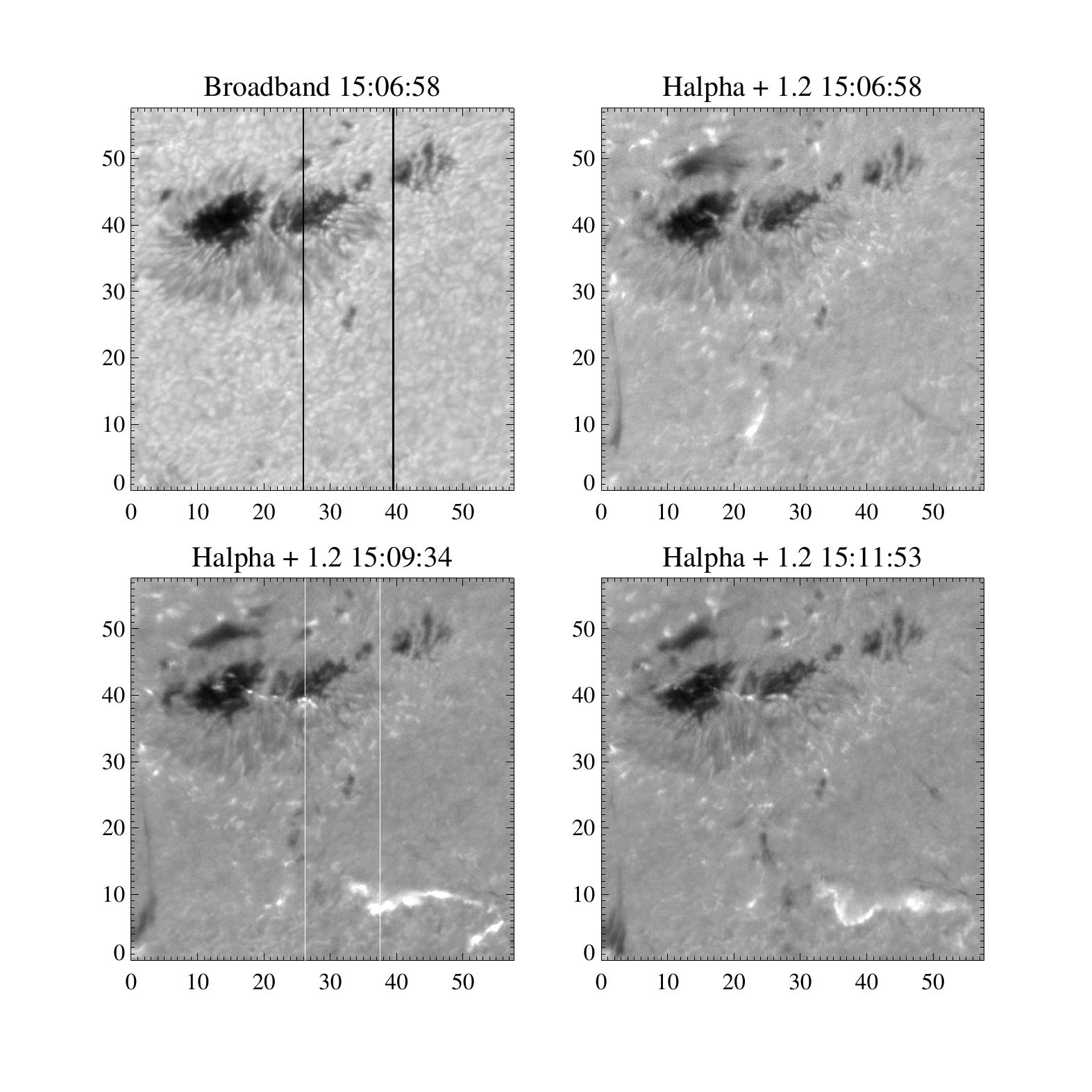}
\caption{ \emph{Abridged caption:}
Central portion of the IBIS field of view. Axes are in units of arcseconds. {\it Top left:} Pre-flare broadband image at 6360 \AA. The two vertical black lines outline the edges of the HSG raster. {\it Top right:} H$\alpha$ + 1.2 \AA, at the same preflare time. An early brightening is already visible within the larger spot. {\it Bottom left:} H$\alpha$ + 1.2 \AA~ near the time of largest hard X-ray peak in the Fermi curve. The left white line indicates the HSG slit position at 15:09:30 and the right white line indicates the HSG slit position at 15:08:44. Excess continuum was detected in the small flare kernel crossed by the slit around position (26\arcsec, 39\arcsec) at this time.  {\it Bottom right:} H$\alpha$ + 1.2 \AA~ at a later time during the flare development.  Note the motion of the plage flare ribbon away from the earlier position.  
The images maintain the native orientation, with vertical direction along the parallactic angle; the east limb direction is roughly towards the bottom of the figure and north is to the right.}
\label{fig:fov}
\end{figure}

IBIS imaged a field of view (FOV) of 98\arcsec\ diameter with a 0\arcsec.098 pixel size, sampling the line
profiles of Fe I 5434 \AA\ (26 steps), H$\alpha$ (30 steps), and Ca II
8542 \AA\ (29 steps). The cadence for the full spectral sequence was $\sim$17~s. We use here
mostly the images acquired in 
the red wing of H$\alpha$ to examine the flare kernel
development.  
Figure \ref{fig:fov} gives an overview of the region and the flare as observed with IBIS.
The two small spots seen in the broadband image as sharing a penumbra were of the same polarity as the leading spot (not in the field of view), and coalesced and grew over the course of two days in the central portion of the active region. This created a compact magnetic neutral line against more sparse plage elements of the following polarity, barely noticeable in the bottom part of the broadband image as bright small features. The co-temporal H$\alpha$$+1.2$  \AA\ image (top right) shows these plage elements much more clearly than in the broadband, due to the relative lack of contrast of convective features at this wavelength, combined with the enhanced temperature of magnetic elements in the mid photosphere  \citep{2006A&A...452L..15L}. 

The two bottom images clearly show the flare ribbons. Flare emission in the far red wing of H$\alpha$ usually displays a very impulsive character and strong spatial and temporal correlation with hard X-ray bursts \citep{1988PASJ...40..357K,1995A&A...299..611C,2011A&A...535A.123R,2013ApJ...769..112D}. Such characteristics are attributed to both local heating and, especially, to the downward moving front of the chromospheric condensation, driven in turn by intense, localized heating such as would be caused by electron precipitation \citep[e.g.][]{1984SoPh...93..105I,1990ApJ...363..318C}. For this reason,  the position of the flare ribbons (or kernels) as imaged in such wavelengths has often been used to identify the electron precipitation site. The bottom panels of Figure \ref{fig:fov}  clearly display the motion of the plage flare ribbon, which proceeds further into the weak field region as the flare progresses, tracing the successive involvement of magnetic field during the flare  \citep[e.g.][]{1997A&A...328..371F}. On the contrary, the spot ribbon does not display any lateral displacement, but rather a succession of bright kernels along a very defined direction (with some repeated episodes in the same kernels). This is most likely due to a very strongly convergent field joining the weak plage to the spot, and determines the extremely narrow ribbon width,  which in various positions approaches the diffraction limit of our observations, $\sim$200 km. Such a property has been observed in other events (always involving a spot ribbon), and can provide important constraints to the standard thick target beam interpretation of solar flares \citep{2011ApJ...739...96K, Sharykin}.

\subsection{Blue/Optical Spectroscopic Data} \label{sec:hsgsetup}
The goal of the optical spectroscopy was to obtain spectra with maximum wavelength
coverage while including the Balmer edge wavelength at $\lambda=3646$ \AA.  
To achieve this, we employed a customized setup of the HSG on the DST.  The solar
spectrum from 3500\,--\,4560 \AA~was imaged over four CCD's in order from bluest to reddest, respectively: ccc06, ccc01, ccc07, ccc08 with instrumental parameters given in Table \ref{table:obsdetails}.  However, only a portion of the spectral range within each CCD was useable, as will be described in Section \ref{sec:reduction}.  A rastering slit with dimensions of 170\arcsec x 0\arcsec.67
was employed with 20 slit positions.  The leftmost slit position in the field of view of Figure \ref{fig:fov} is referred to as the first slit position throughout the paper.  The step-size between each slit position was
0\arcsec .6775, giving a total raster extent of 170\arcsec $\times$13\arcsec .5, and the total cycle time was $\sim21$~s.  
The spectra were obtained with the slit oriented perpendicular to the
horizon (at the parallactic angle), to ensure that image displacements due to differential refraction would align with the spectrograph slit. Such displacements are clearly noticed in high spatial resolution solar observations  \citep{2006SoPh..239..503R}, and are of the greatest relevance at the blue wavelengths we employ in this study. The raster direction was perpendicular to the slit; this made for an orientation of the field of view in Figure \ref{fig:fov}, Figure \ref{fig:raster2}, and Figure \ref{fig:rasterHa} different from the standard orientation with solar North up. 

We also obtained high spatial sampling images
($\sim$0\arcsec .075 pixel$^{-1}$) with the slit jaw camera
through the NBF4170 filter, typically employed with the ROSA
instrument \citep{Jess2010}.  This filter is centered on
$\lambda=4170$ \AA\ with a bandpass of 52 \AA\ and an exposure time of
10 ms.  The slit jaw images allowed us to accurately determine the position of the slit on
the Sun.

\subsection{Spectroscopic Data Reduction} \label{sec:reduction}
The spectroscopic data reduction was performed using standard
IRAF\footnote{IRAF is distributed by the National Optical Astronomy
  Observatory, which is operated by the Association of Universities
  for Research in Astronomy (AURA) under cooperative agreement with
  the National Science Foundation.} and IDL routines.  We corrected
all images for dark current.  De-focused quiescent solar spectra obtained away from the active
region were used for flat field, wavelength, and intensity calibration, which is described in detail 
in Appendix A.  The reference spectrum for calibration was the disk-center absolute solar intensity spectrum obtained with the Fourier Transform Spectrometer (FTS) with spectral resolution R = 350\,000 \citep{Neckel1999}.  
The nominal dispersions for each camera are given in Table \ref{table:obsdetails}, but 
we note that, as a compromise between spatial scale along the slit and exposure times, the slit width was fixed at 90 $\mu$m, corresponding to an actual
spectral resolution of 0.9\,--\,1.2 \AA\ at 4300 \AA~
(R$\sim$4000).  We converted the 2D spectra to intensity ($I_{\lambda,\mu=0.74}$; ergs cm$^{-2}$ s$^{-1}$ sr$^{-1}$ \AA$^{-1}$) by accounting for limb darkening, instrumental sensitivity, and the atmospheric extinction. The spectra from each CCD were aligned and
interpolated to a common pixel scale (0\arcsec.39 pixel$^{-1}$).  Wavelength-dependent shifts of $0.5 - 2$ pixels were applied to account for differential refraction within each camera's spectral range.  

As demonstrated in Figure \ref{fig:calibration}, we found a satisfactory agreement between the observed quiescent 
spectrum (obtained at approximately $(x, y)=$ (38\arcsec, 20\arcsec) in Figure \ref{fig:fov}; i.e., between the plage regions)
 and the disk-center reference FTS spectrum binned to the HSG spectral resolution and converted to $I_{\lambda, \mu=0.74}$.  
Unfortunately, the current DST optical path is not optimized for work over very broad wavelength ranges, and suffers
from chromatic aberration.
The bluemost camera suffered
the most from this problem, displaying differential spatial and spectral focus in large portions of its
range, except for pristine focus in the interval
$\lambda=3654-3674$~\AA.  Interpretation of the data outside this spectral region 
is especially problematic in the umbral regions where there are sharp 
spatial gradients in intensity.  At $\lambda < 3600$~\AA\ and $\lambda >
3740$~\AA, the chromatism becomes severe and spectral and spatial
features are largely defocused.  In the figures, we show the spectral
range from $\lambda=3600-3740$~\AA; although characterization is not
robust through this entire spectral range, the degree to which we can 
detect the continuum and line features is satisfactory.  Besides these problems, 
the chromatism was evident in the ccc07 
camera from 4016\,--\,4200 \AA, such that solar features 
at $\lambda=4200$ \AA\ were sharper than the features at
$\lambda=4016$ \AA.   The focus differs slightly among the four CCD's, with 
excellent overall focus in ccc08, and poor overall focus in ccc01. 

Based on comparisons of our data to the disk-center FTS spectrum, we
found the useable wavelength ranges are the following:
3654\,--\,3674 \AA\ for ccc06,
3830\,--\,3978 \AA\ for ccc01, 4085\,--\,4125 \AA\ for ccc07, and
4213\,--\,4553 \AA\ for ccc08. 

\section{White-Light Detection} \label{sec:WLresults}
Historically, many different parameters have been used to characterize white-light emission in solar flares.  \cite{Jess2008} has demonstrated that 
ambiguity can result if the quantities are not precisely defined.  Here, we calculate the enhancement, excess intensity, and contrast, in order to allow
a meaningful comparison to the variety of measurements in older literature.

\subsection{Continuum enhancement} \label{sec:WL_enhancement}
The \emph{enhancement} (or ratio) images are obtained by dividing the intensity at the time of the flare by the pre-flare intensity at UT 15:07:24 at the same spatial location.  Figure \ref{fig:raster2} shows enhancement images for the continuum ($\lambda \sim 4170$ \AA; top panels) and at H$\delta$ line-center ($\lambda=4101$ \AA; bottom panels). In Figure \ref{fig:rasterHa} we also show a similar field of view extracted from the IBIS H$\alpha+1.2$ \AA\ data at multiple times, which allows the red wing kernels to be compared to the location of the blue/optical enhancements.

\begin{figure}[h]
\centering
\includegraphics[width = 6.5in]{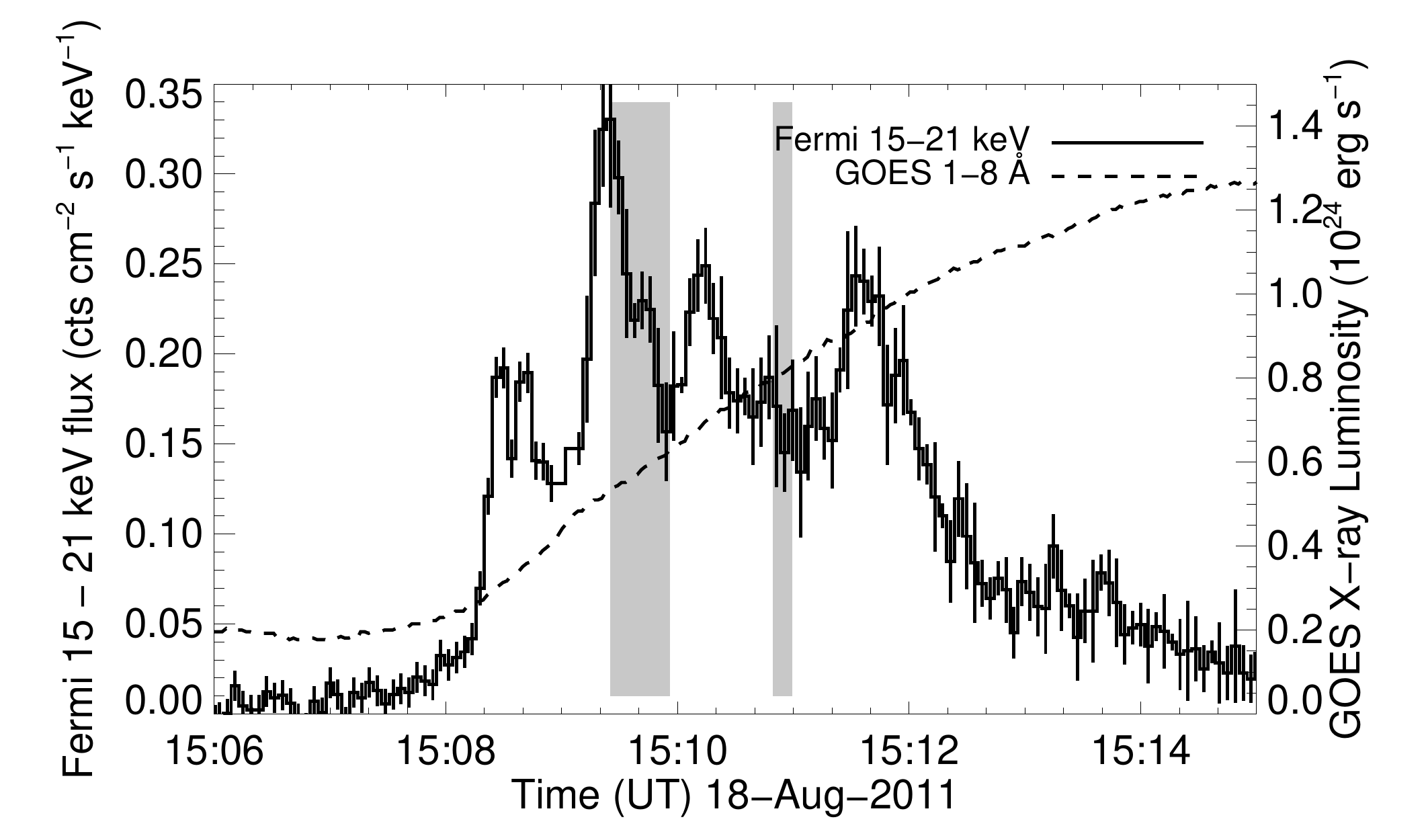}
\caption{ The Fermi $15-21$ keV hard X-ray count flux light curve (left axis) shown with the GOES $1-8$ \AA\ luminosity (right axis) of the C1.1 flare SOL2011-08-18T15:15 from AR 11271.  The timing of the simultaneous H$\alpha+1.2$ \AA\ and 
optical continuum enhancements are indicated by vertical grey bars.  Each Fermi data point has a live time of 4.07~s. }
\label{fig:fermi_goes}
\end{figure}

The enhancement images in the continuum are quite noisy, with many scattered 
small scale features whose intensity changes by a few percent between time steps. This is due mostly to transparency and seeing fluctuations (which have the largest effect in areas of large intensity gradients and cannot be removed from slit spectra), as well as to general evolution of the structures, for example the slow variation of brightness in the plage elements at the bottom of the panels. However, the very bright small feature at the leftmost slit position, indicated by the white arrow in the 15:09:30 panel, is a  genuine candidate for a WL enhancement in the umbral region. Indeed, this feature appears at the same position along the slit as the umbral flare kernel shown in the 15:09:34 panel of Figure \ref{fig:fov} and, most importantly, its temporal evolution follows closely that of the umbral flare kernels as observed both in the core of H$\delta$ (bottom panels of Figure \ref{fig:raster2}) and in the wing of H$\alpha$ (Figure \ref{fig:rasterHa}).  In particular, the WL brightening is readily discernible at the same spatial location in three consecutive images, 15:09:08, 15:09:30 and 15:09:51, reaching an enhancement $\ge$ 1.25 at 15:09:51 UT. After fading away, it re-appears briefly about 1 min later, at 15:10:54, again with an enhancement of $\sim$1.25, consistent with the repeated appearance of the umbral kernel in the H$\alpha$ wing images at 15:11 (Figure \ref{fig:rasterHa}). 
Figure \ref{fig:fermi_goes} shows the Fermi hard X-ray ($15-21$ keV) and GOES soft X-ray ($1-8$ \AA) light curves with the times of the simultaneous umbral kernel enhancements observed in the blue/optical continuum and at H$\alpha+1.2$ \AA\ as grey vertical bars.  This shows that the kernel ``flickering'' occurs in the impulsive phase well before 
the maximum soft X-ray emission and is generally consistent with the well-established temporal
coincidence of hard X-ray and the white-light emission \citep{Kane1985, Hudson1992, NeidigKane1993, Fletcher2007}.  

As commented on in Section \ref{sec:IBIS}, the spatial extent of the flare umbral kernels is very small. In particular, the kernel intersected by the slit at 15:09:08 is roughly circular in shape, with a diameter of 5 IBIS pixels, i.e. $\le$ 0\arcsec.5. This is consistent with the size of the WL enhancement shown in Figure \ref{fig:raster2}, which extends to at most two pixels along the slit, while being detectable at a significant level only in the first slit position of the raster.  Thus, the WL kernel appears spatially unresolved in the HSG data.  Using the circular figure from the IBIS data, we find an upper limit for the area encompassing the H$\alpha$ wing excess intensity of $\sim10^{15}$ cm$^{2}$ (1\arcsec $=$ 734 km).  This is comparable to the areal coverage of the WL kernel observed by \cite{Jess2008} during a C2.0 flare, at $\lambda=3953.7\pm5$ \AA.  In Section \ref{sec:discussion}, we show that the resolved area is important for estimating the actual intensity values.  

\subsection{Excess  $\left< I_{\lambda} \right>$} \label{sec:WL_excess}
The \emph{excess spatially averaged intensity}, or $excess \left< I_{\lambda} \right>$, is defined as 
the pre-flare intensity at 15:07:24 subtracted from the flare intensity and averaged over the 
spatial extent of 3 pixels (1\arcsec.2$\times$0\arcsec.67) centered on the brightest pixel.  The averaging was done
to account for slight residual spatial misalignments of the four CCDs and to account for the PSF of the 
slit.  
We note that, although the excess intensity can be used directly for calculating the flare energetics (see Section \ref{sec:discussion}), it
 does not directly relate to an intensity from a given emission mechanism unless the emission is completely optically thin\footnote{As noted by \cite{Acampa1982}, the excess emission is a metrological quantity; see also the discussion in \cite{KerrFletcher}. }. It is however  
useful for detecting a low level of flare emission.  

In Figure \ref{fig:cont_evol}, we show the time-evolution of the
excess emission in continuum regions within each camera at the
umbral kernel position.
Note that these are lower limits to the excess intensity
since the umbral kernel
is unresolved in the spectra (but see Section \ref{sec:discussion}). 
The two episodes of statistically significant continuum brightenings at 15:09:30\,--\,15:09:51 and at
15:10:54 are highlighted with grey bars in Figure \ref{fig:cont_evol};
the continuum excess is evident across the full spectral range.  
Histograms of the excess intensity value per pixel at 15:09:30 at selected continuum wavelength regions and in H$\gamma$ are shown in
Figure \ref{fig:all_resid}, demonstrating that the intensity in the umbral kernel 
is well outside the spatial fluctuations in the data (e.g., top panel of Figure \ref{fig:raster2}).  The intensity values in the three pixels that are averaged for the umbral kernel are indicated by vertical dotted 
lines; at least one of each set of three pixels is $\ge 3\sigma$ of the
distribution.

As mentioned, atmospheric seeing variations
 can induce fluctuations in intensity over time, with particularly strong effects near large
gradients in intensity, such as within the umbra.  In the bottom panel
of Figure \ref{fig:cont_evol}, we show the average excess intensity
variations from a non-flaring umbral region with an average intensity
level and gradient similar to that of the umbral kernel. 
The standard deviation of the
light curve of the non-flaring umbral excess gives
an estimate of the statistical fluctuation; we adopt this variation for the error bars in the 
top panel of Figure \ref{fig:cont_evol}. The excess values within the time ranges indicated by the 
grey bars in the top panel of Figure \ref{fig:cont_evol} occur at a confidence level of $4.5-5 \sigma$ in the four
continuum regions.  
The continuum variations in this location during times outside the grey bars are not
significant enough to conclude they are true enhancements.  

Any similar continuum excess outside the umbral region (at other slit positions) would give a lower enhancement, and may
not be detectable from visual inspection of Figure \ref{fig:raster2}.  Therefore, we performed a systematic
search over all times and all slit positions, requiring that the excess H$\gamma$ and the excess continuum intensity at $\lambda=3654-3674$ \AA\ exceed a significance of 5$\sigma$ and 3$\sigma$, respectively, where $\sigma$ is determined from the spatial variation of the excess along the slit at each slit position (as in Figure \ref{fig:all_resid}).  In addition to the umbral kernel, we find a candidate continuum increase at 15:09:47 in the 19th slit position in the plage ribbon, coincidentally within the time range of the umbral enhancement.  However, this signal is only significant at a 3$\sigma$ level in the bluemost camera and 2$\sigma$ in the other cameras.  At this level, we cannot conclude that this is a bona-fide WL excess.

\begin{figure}[h]
\centering
\includegraphics[width = 6.5in]{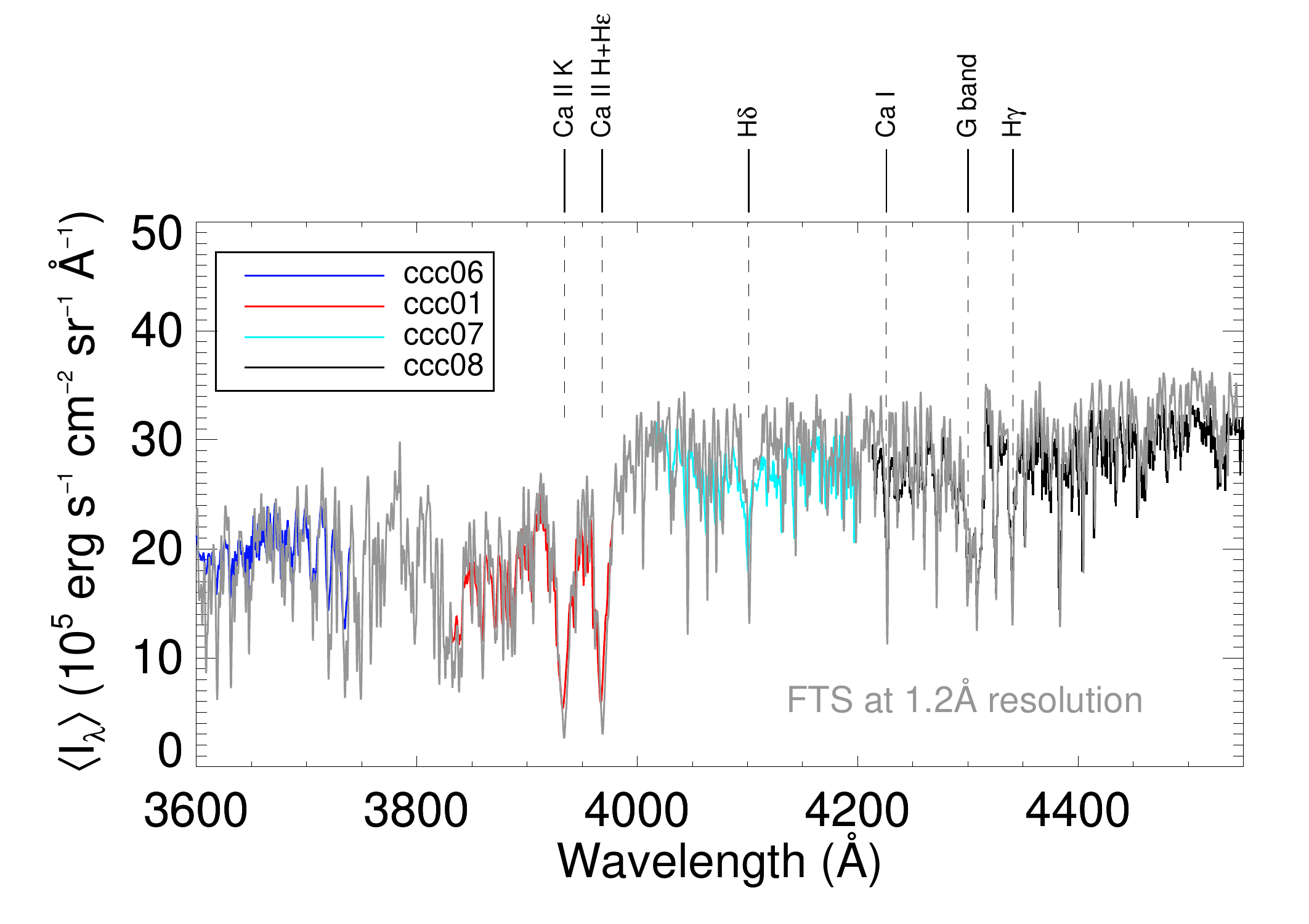}
\caption{Total intensity (averaged over 3 spatial pixels; $\left<I_{\lambda}\right>$) of a non-flaring granulation region away from plage 
and umbra at the 18th slit position at 15:07:20, compared to the FTS disk-center
intensity adjusted by the limb darkening at $\mu=0.74$ and convolved with a Gaussian of FWHM$=1.2$ \AA.   The intensity level and shape of the observed solar continuum is reproduced well in this quiescent region.
}
\label{fig:calibration}
\end{figure}

\begin{figure}[h]
\centering
\includegraphics[width = 6in]{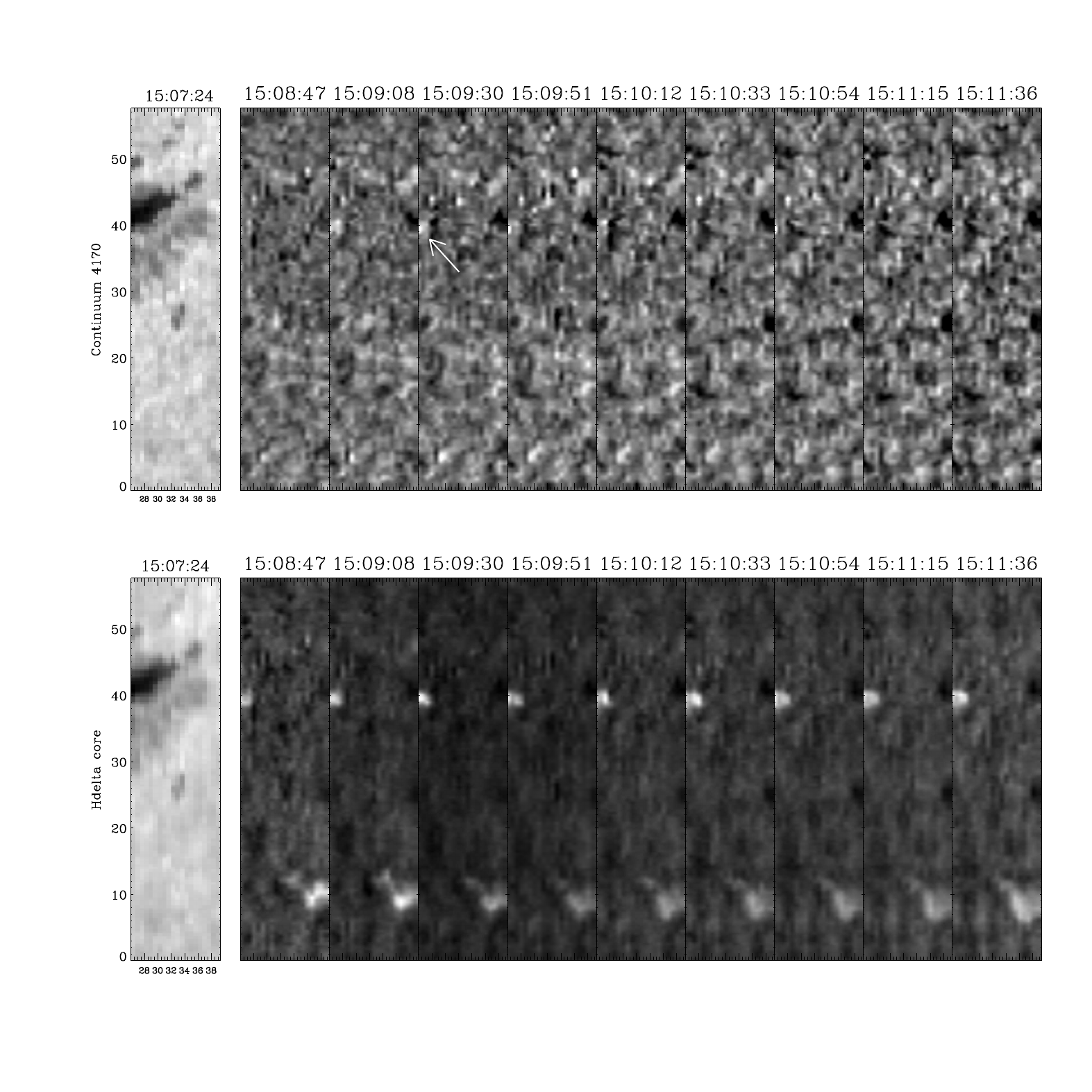}
\caption{{\it Top panels:} HSG raster continuum image in the pre-flare (left panel), and images of the continuum ratio (enhancement) at 4170 \AA~at various times during the flare development. The enhancement images are scaled between $-$10\% and $+$15\% of the pre-flare values in the same spatial positions. The white arrow indicates the WL enhancement discussed in the text. {\it Bottom panels:} the same as top panels, for the H$\delta$ line core, scaled between $-$30\% and $+$40\%. Note the sharper definition of the continuum image, highlighting photospheric features, and the larger extension of the flare kernels as imaged by the chromospheric line core emission. The time indicated above the panels refers to the beginning of each raster scan, which proceeds from left to right in the images.}
\label{fig:raster2}
\end{figure}

\begin{figure}[h]
\centering
\includegraphics[width = 6.5in]{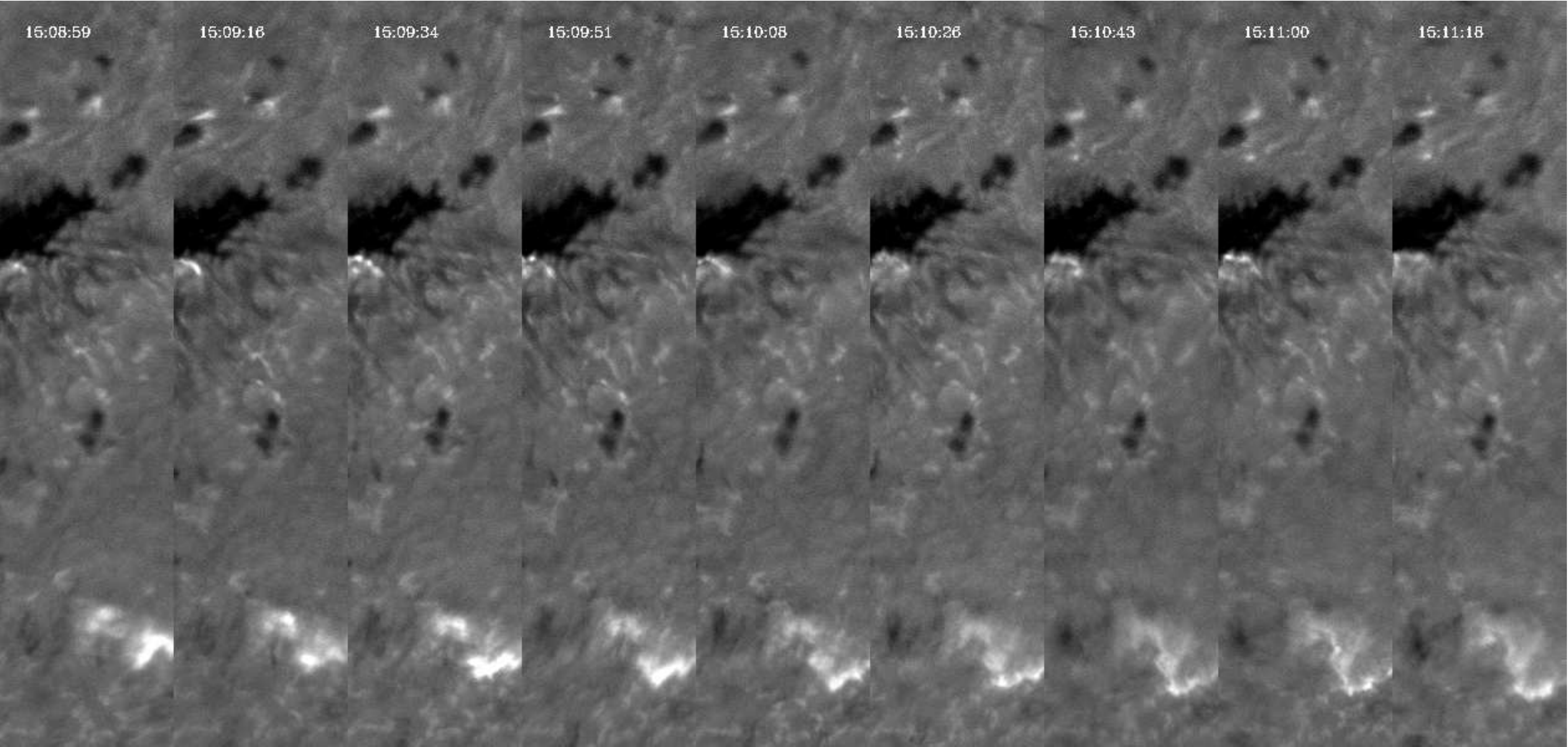}
\caption{IBIS data showing the H$\alpha+1.2$ \AA\ wing evolution for the approximate field of view covered by the 
HSG spectral raster. }
\label{fig:rasterHa}
\end{figure}

\begin{figure}[h]
\centering
\includegraphics[width = 6.5in]{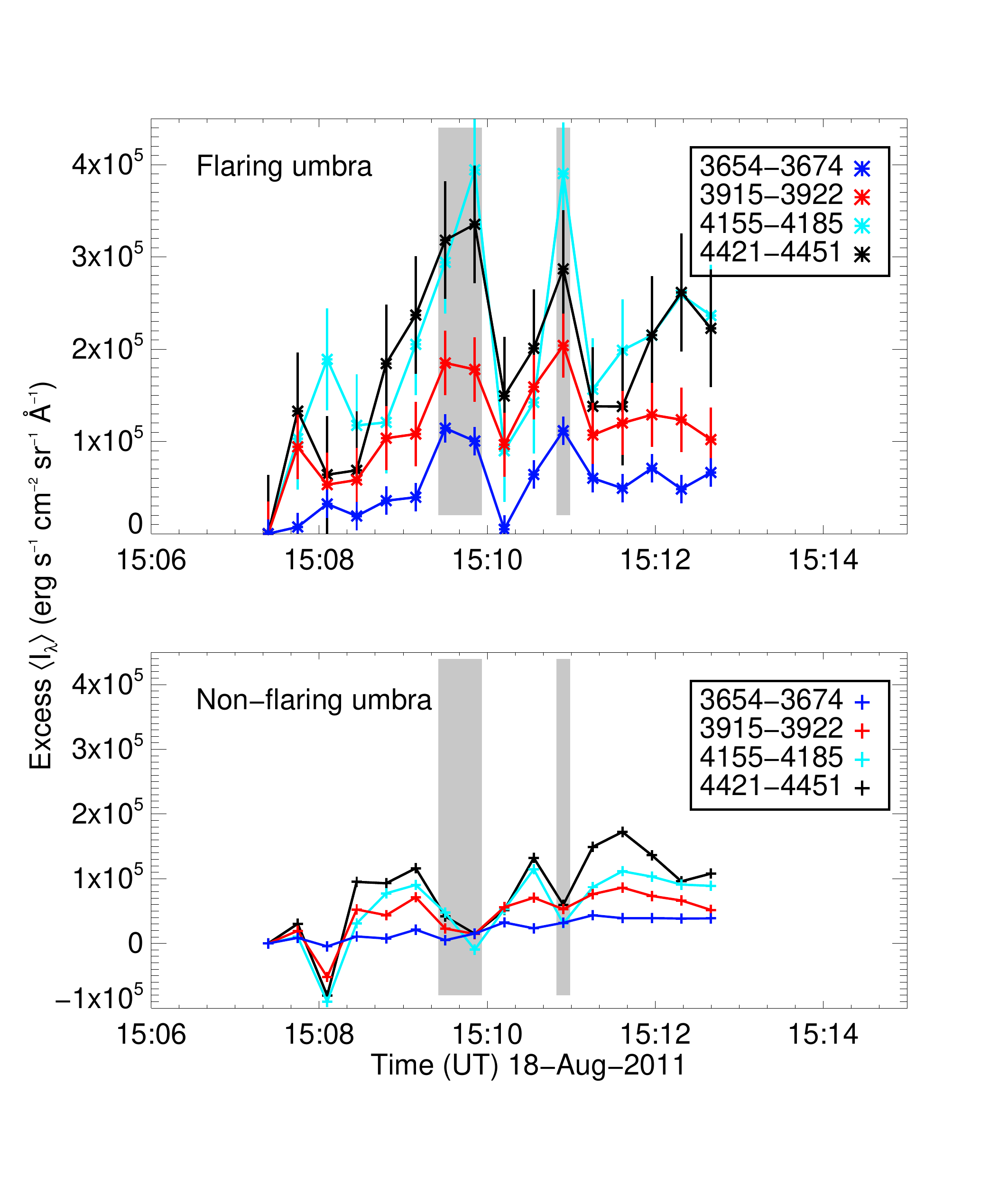}
\caption{\emph{Top:} The time-evolution of the excess continuum intensity in four
  spectral regions, extracted from the umbral kernel. 
  The vertical
  grey bars indicate the times of the significant flare continuum detections. \emph{Bottom:} The
  excess continuum variations in a nearby non-flaring region of the
  umbra.  The standard deviation of this panel gives the statistical
  error in the top panel light curve. }
\label{fig:cont_evol}
\end{figure}

\begin{figure}[h]
\centering
\includegraphics[width = 4.5in]{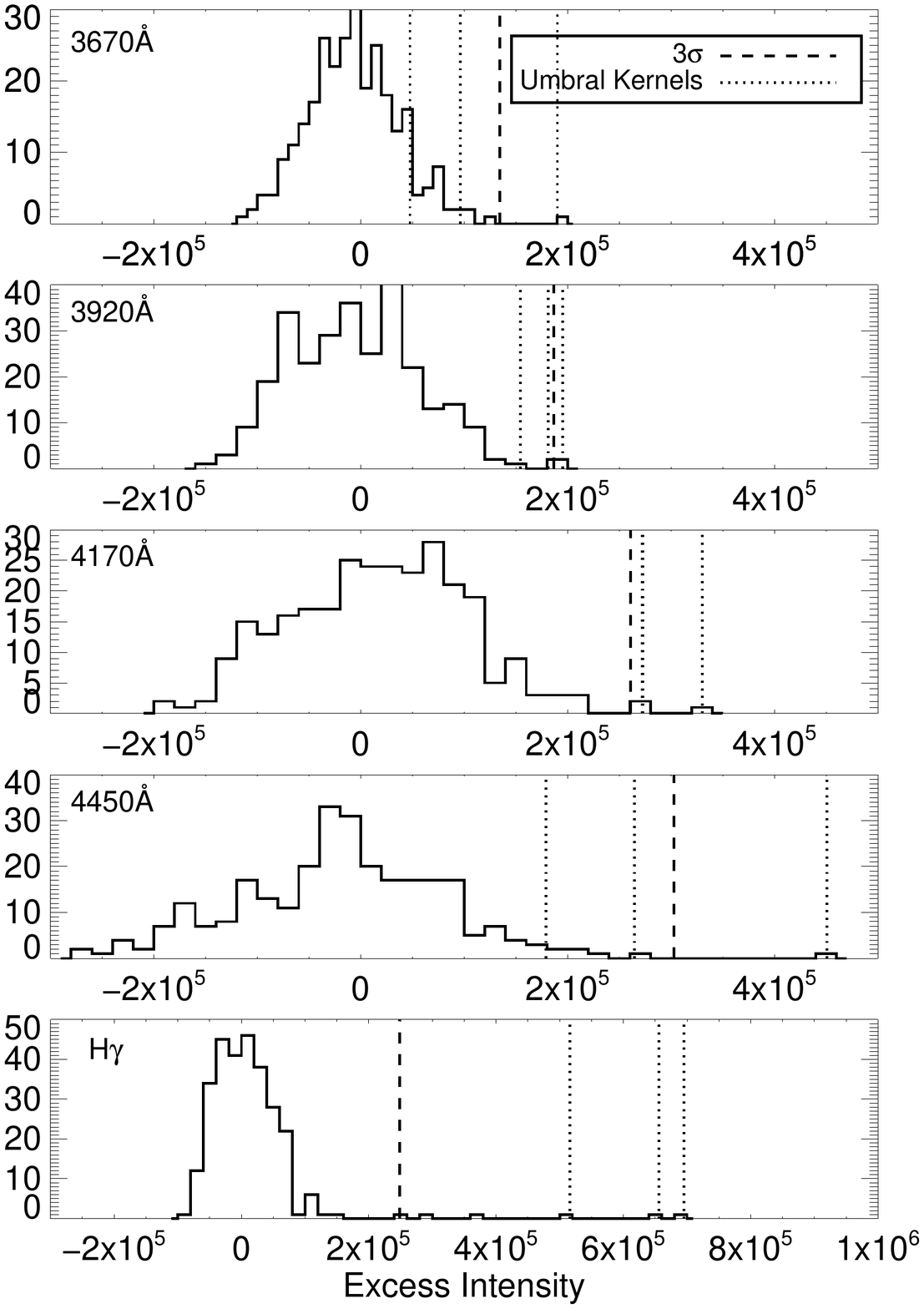}
\caption{Histograms of the excess intensity (subtracting 15:07:24 from
15:09:30) over a spatial cut through the HSG data, at selected
continuum intervals and H$\gamma$.  The value corresponding to
3$\sigma$ of the distribution is indicated by dashed lines, whereas the values
for the 3 pixels averaged to obtain the excess $\left<I_{\lambda}\right>$ for the umbral kernel are indicated by dotted lines.
Note that the value of $\sigma$ in this figure represents the spatial variation of the excess, whereas the value of $\sigma$ in Figure \ref{fig:cont_evol} describes the temporal variation of the excess. }
\label{fig:all_resid}
\end{figure}

\begin{figure}[h]
\centering
\includegraphics[width = 5.0in]{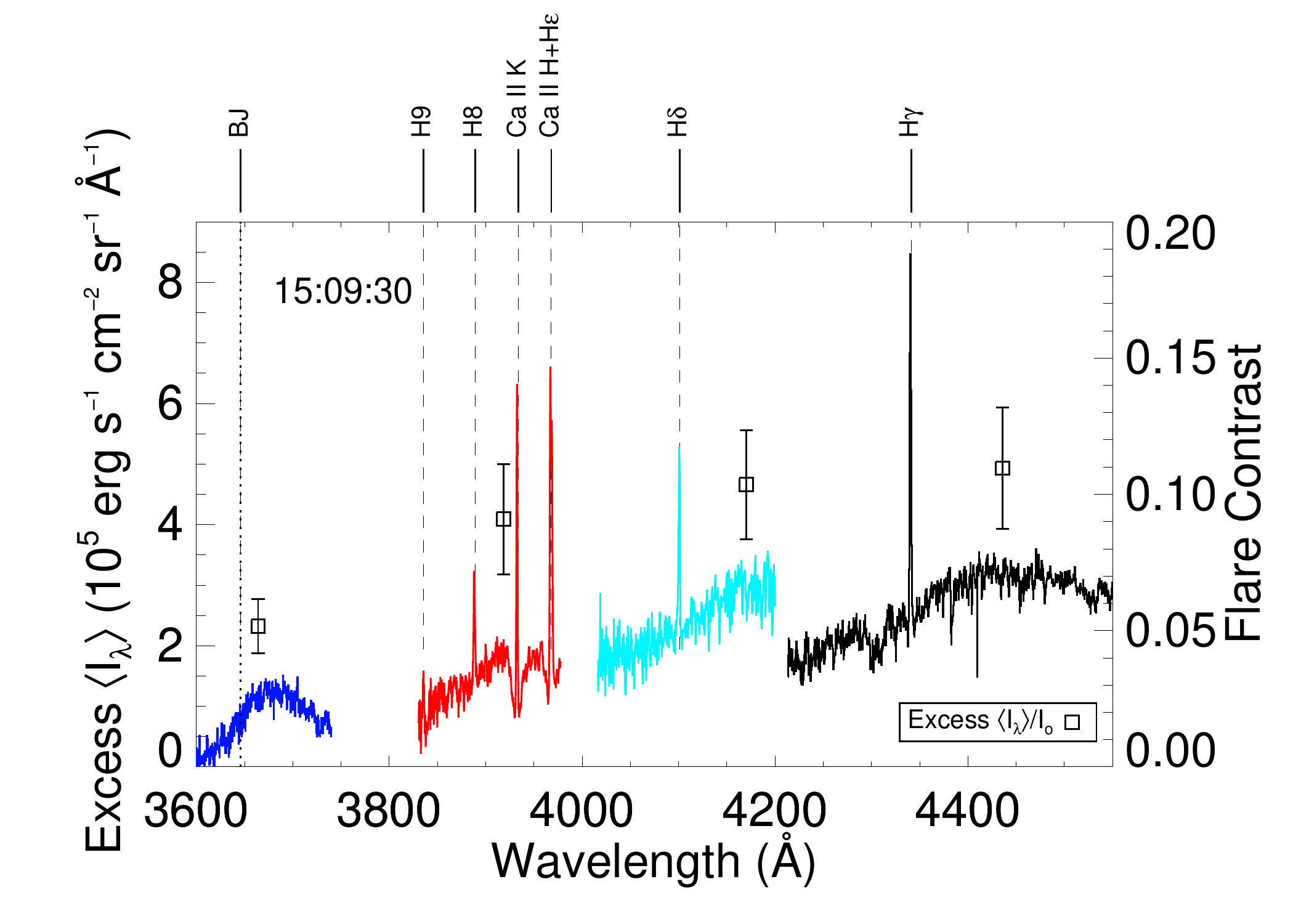}
\caption{The full spectral range of the excess intensity at 15:09:30 in
 the umbral kernel  at the first slit position in the spectral raster.  
  The intensity is averaged over three spatial pixels.  The right axis (square symbols)
   show the flare contrast in selected continuum wavelength regions. }
\label{fig:spectrum1}
\end{figure}

Figure \ref{fig:spectrum1} shows the full spectral range of the spatially averaged excess
intensity at 15:09:30 in the first raster slit 
position for the umbral kernel.
The umbral kernel spectra at 15:09:51 and 15:10:54 exhibit similar 
continuum properties and are not shown.  
The rising intensity from $\lambda=4000$ \AA\ to $\lambda=4200$ \AA\ is probably a residual effect of the strong intensity gradient experienced by this camera  from blue to red wavelengths (due to both the solar spectrum and the detector spectral response), made evident because of problems with non-linearity at the low counts of the sunspot spectrum. Also, the strong chromatic aberration across this chip (see Section
\ref{sec:reduction}) might cause the mixing of adjacent spectra from vastly different features, a problem of further relevance in the case of strong spatial intensity gradients as in the spot-penumbra transition.  
This likely results in the apparent continuum jump at $\lambda=4200$ \AA\ in Figure \ref{fig:spectrum1} (between the two red most cameras), preventing a detailed slope characterization, and we display it only for the purpose of continuum detection.  A peculiar continuum feature is the ``bump'' between $\lambda=4400$ \AA\ and
$\lambda=4500$ \AA.  \cite{Donati1984} also observed a continuum bump
near this wavelength in their solar flare spectra, which for now remains unexplained.

The excess intensity in the Balmer jump region in this umbral kernel is of particular
interest for modeling constraints.  In Figure \ref{fig:balmerjump}, we
show the excess intensity at 15:09:30, 15:09:51, and 15:10:54 in the
bluemost spectral region.  The excess has little variation among the
three times.
As mentioned before, pristine focus is only achieved from
$3654-3674$ \AA, which is indicated by the shaded region in Figure \ref{fig:balmerjump}.  Therefore, the
slopes outside of this range cannot be characterized with high
significance, due to the ambiguity from subtraction of the pre-flare
(umbral) intensity level, which has a drastic spatial variation.  
Despite this, a broad continuum
feature redward of the predicted Balmer edge
at $\lambda = 3646$ \AA\ is well noticeable.  The excess intensity extends from $\lambda 
\sim3646$ \AA\ while apparently increasing towards a broad maximum
centered roughly at $\lambda \sim 3675$ \AA, which has an excess average intensity of $\sim$1.2$\times10^{5}$ erg s$^{-1}$ cm$^{-2}$ sr$^{-1}$ \AA$^{-1}$.  We discuss this feature further in Section \ref{sec:discussion} in relation to previous studies.
In the figure, the expected wavelengths of the higher order
Balmer lines (H13-H19) are also indicated.  We can identify Balmer lines in emission
up to H14.  In our spectra, an emission line is clearly located near
the standard wavelength of H16 ($\lambda3704$), but He \textsc{i}
($\lambda$3705.0) and Fe \textsc{i} ($\lambda3705.6$) have been 
observed with just as large or larger flux as H16 in spectra of stellar
flares \citep{Hawley1991, Fuhrmeister2008}.  

Finally, we measure the ratio of excess
continuum intensity in the bluest camera to the excess at the selected
continuum regions at redder wavelengths at $\lambda=3915-3922$ and at $\lambda=4421-4451$ \AA\ from Figure \ref{fig:cont_evol}) for comparison to model values in a future paper.  These spectral
ranges are selected where focus is best and chromatic aberration does not affect the intensity. 
The ratios of excess continuum intensity at 15:09:30 are $\sim$0.6 and 0.4, respectively.

\subsection{Flare contrast} \label{sec:WL_contrast}

\begin{figure}[h]
\centering
\includegraphics[width = 5.0in]{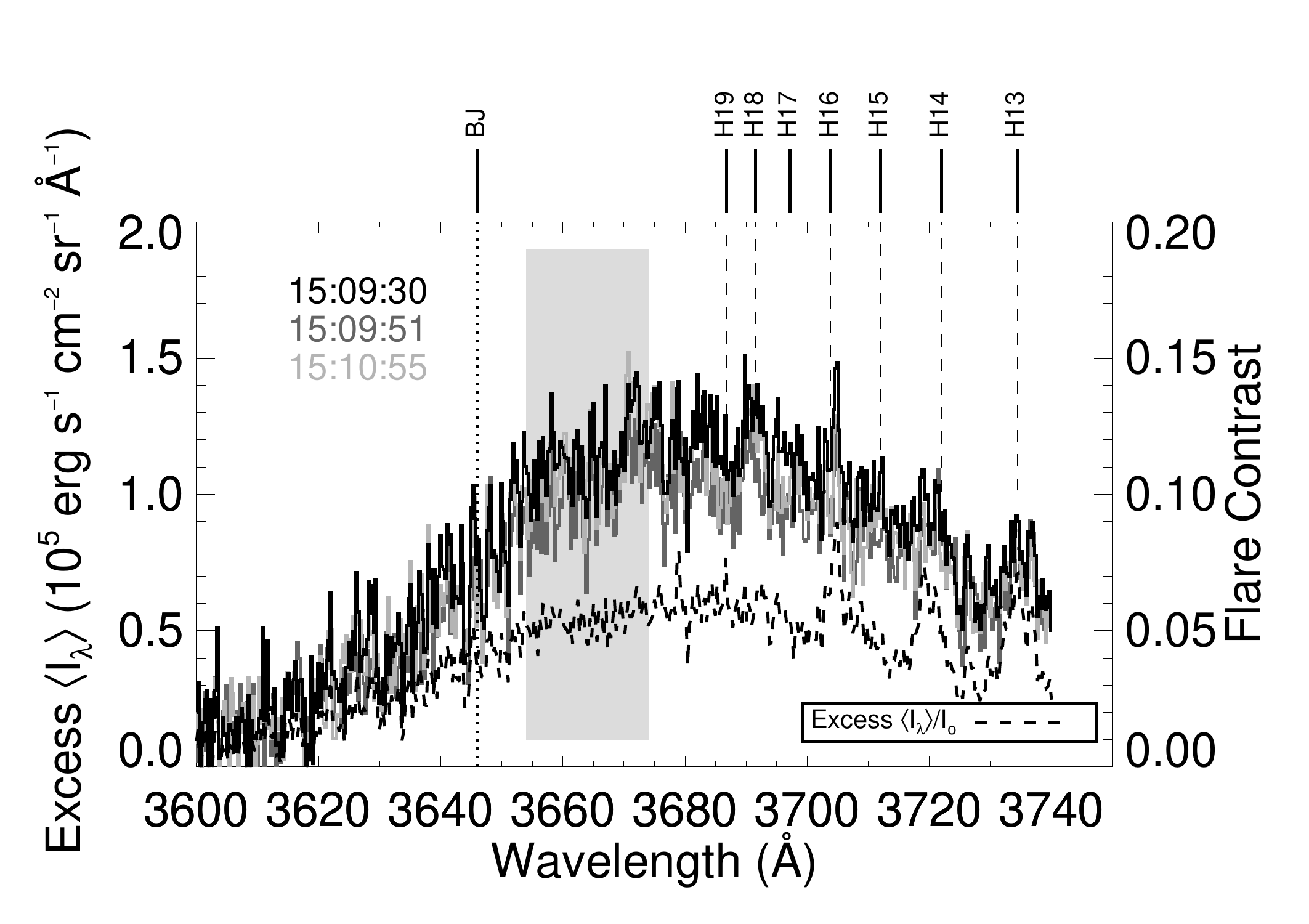}
\caption{The excess intensity in the
  umbral kernel 
  at the first slit position in the spectral raster,
  shown for the bluemost spectral region at three times.
The expected wavelengths of the Balmer jump (BJ) and the higher order Balmer lines are indicated; H14 is the last visible
 Balmer line, and the feature near the wavelength of H16 could be a blend of Fe \textsc{i} and He \textsc{i} at this 
wavelength (see text).  The shaded area indicates the wavelength region not affected by chromatic aberration.
  The right axis (dashed line)
   shows the flare contrast.  }
\label{fig:balmerjump}
\end{figure}

An additional parameter used to characterize white-light emission in flares is the  \emph{flare contrast}, or $excess \left<I_{\lambda}\right>$/$I_o$ where $I_o=I_{\rm{granulation}}$ is the non-flaring solar granulation intensity in Figure \ref{fig:calibration}. To facilitate comparison with earlier spectra \citep[e.g.,][]{Neidig1983}, 
we show in Figure \ref{fig:spectrum1}  (square symbols, scale on the right axis) a measure of the flare contrast for the four continuum windows from Figure \ref{fig:cont_evol}.  
The flare contrast is $\sim$10\% throughout the spectral range, with slightly lower contrast of $\sim$5\% in the far blue just redward of the Balmer edge wavelength.  
Note, some previous measurements of flare contrast allowed the subtraction of 
a nearby spectrum of the quiet sun at the same time as the flare.  In our flare, the total intensity is low compared to granulation, so  
we must subtract the pre-flare umbral region to obtain a meaningful (positive) quantity. 
The flare contrast at 15:09:30 
is also indicated in Figure \ref{fig:balmerjump}.
It exhibits a similar trend to the excess.

\section{Emission Line Analysis} \label{sec:timing}

In addition to the significant continuum enhancement,
several chromospheric emission lines are present in the flare spectra:
H$\gamma$ ($\lambda4341$), H$\delta$ ($\lambda4101$), Ca
\textsc{ii} H (blended with H$\epsilon$, $\lambda3968$), 
Ca \textsc{ii} K ($\lambda3934$), H8 ($\lambda3889$), and H9 ($\lambda3835$).  We describe here their properties.

\subsection{Comparison with the Hard X Ray Fermi Light Curve}
\begin{figure}[h]
\centering
\includegraphics[width = 5.0in]{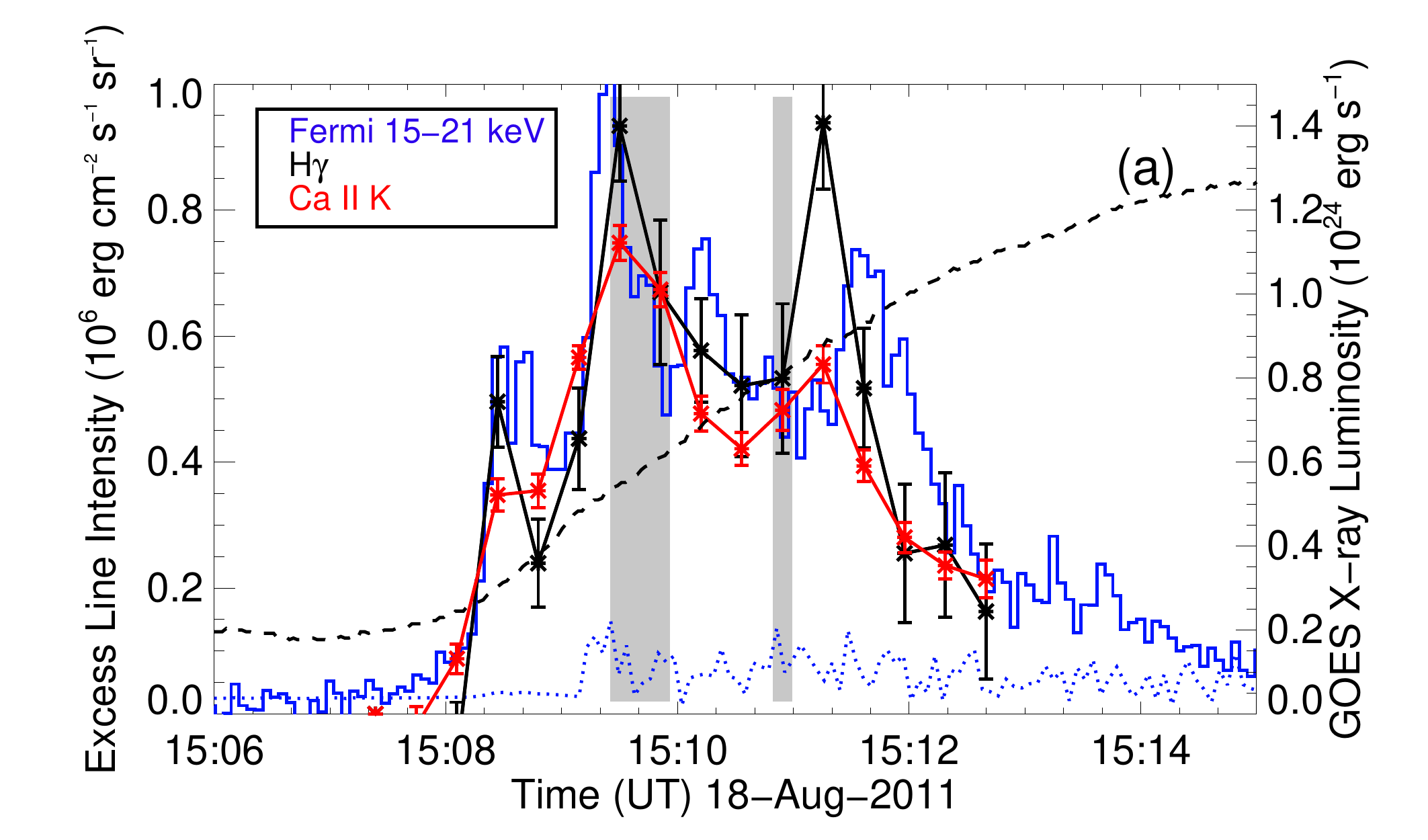}
\includegraphics[width = 5.0in]{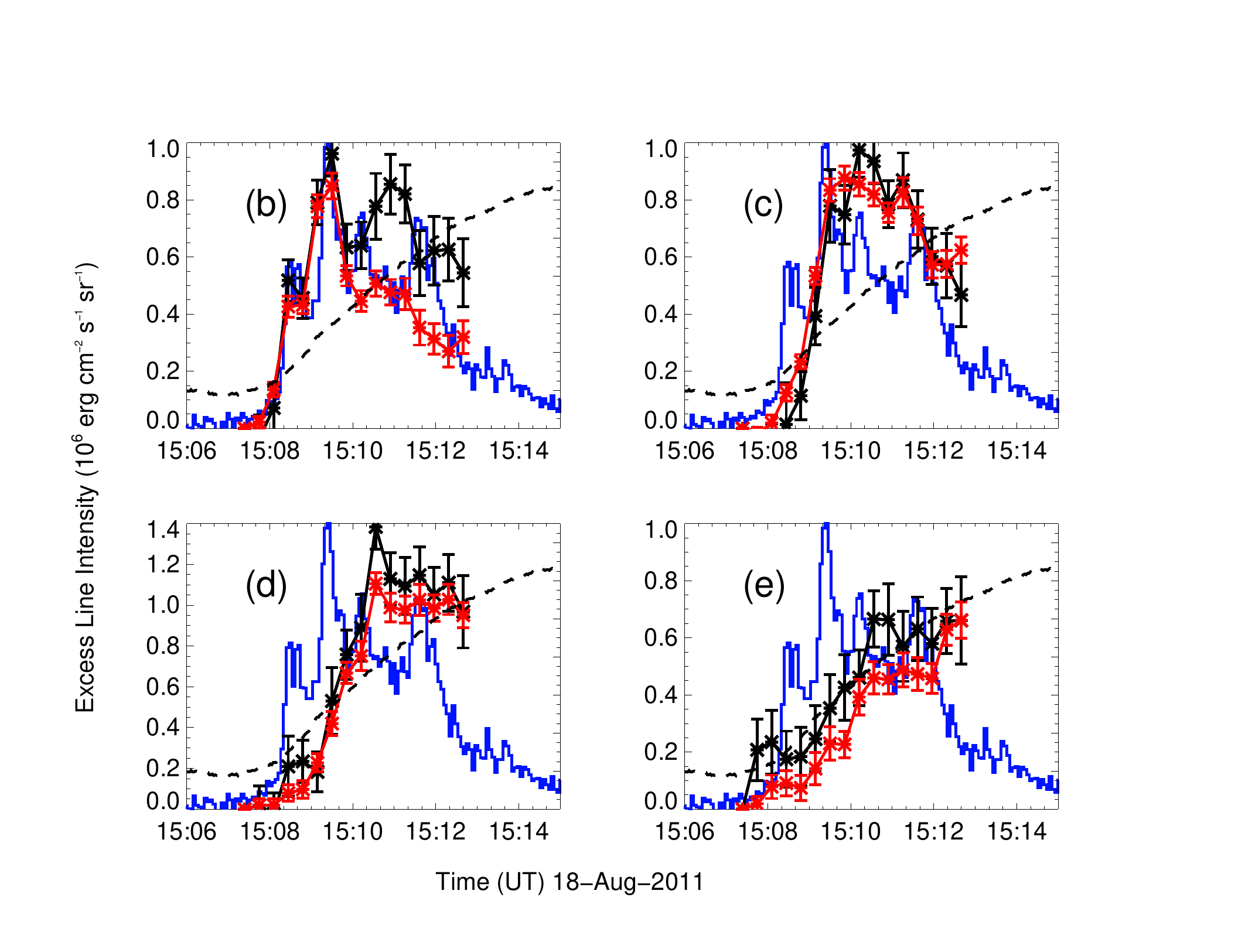}
\caption{(a) The light curves of Ca \textsc{ii} K
and H$\gamma$ emission lines from the umbral kernel compared to the Fermi/GBM
data from 15-21 keV (normalized to the peak value of 0.33 counts cm$^{-2}$ s$^{-1}$ keV$^{-1}$).  
The average excess line intensity over three spatial pixels is shown.  The dashed line is the GOES
1-8 \AA\ luminosity (right axis), and the dotted line is the 1$\sigma$ error for the Fermi data.  
Grey vertical bars indicate the times of 
a significant continuum excess.  Panels (b)-(e) show the same quantities for the regions adjacent to 
the umbral kernel, in the second, third, fourth, and fifth slit positions, respectively.  Note the rescaling of the y-axis in panel (d). }
\label{fig:fermi1}
\end{figure}

In Figure \ref{fig:fermi1} (a-e) we show the continuum-subtracted, line-integrated 
excess intensity as a function of time in H$\gamma$ and Ca \textsc{ii} K for the first slit position at the location
of the umbral kernel (a), and in adjacent regions to the umbral kernel in the second, third, fourth, and fifth
slit positions (b-e, respectively).  The uncertainties of the integrated excess 
line intensity were calculated following the standard formula in the Appendix of \cite{Chab}, which
adds the uncertainty of the integrated continuum and line excesses in quadrature.  
  Although the emission line excess extends over a larger region ($\sim$3\arcsec, Figure \ref{fig:raster2}) compared to the continuum excess, we
average the intensity only over the three brightest pixels along the slit.  The emission line light curves are compared to the
$15-21$ keV hard X-ray light curve obtained from Fermi/GBM and the 1-8 \AA\ soft X-ray luminosity obtained from GOES.  The same grey bars from Figure \ref{fig:cont_evol} indicate the times of significant continuum excess in panel (a).  
In the leftmost slit position of the raster (a), 
we see a general similarity in the normalized time variation of the hard X-rays and the optical
lines.  As we progress away from the leftmost slit position (panels b-e), we observe
a more gradual response in the optical lines, yet reaching a comparable maximum value
as in the first slit position.  By the fifth slit position, i.e. $\sim$2\arcsec.5 apart, the optical lines appear to evolve similarly to the soft X-ray emission.  The gradual evolution in the chromospheric emission lines coincides with the formation of new, low-brightness kernels at these later times, as seen in the development of the H$\alpha+1.2$ \AA\ umbral ribbon (Figure \ref{fig:rasterHa}). 

At the location of the umbral continuum excess, we find coincident peaks between the hard X-ray and the 
optical emission line light curves, but the 20 s cadence of the optical lightcurves makes it difficult to compare the precise timing with the much better sampled hard X-ray light curve ($\Delta t \sim $4~s). 
The first enhancement at 15:08:27 in the hard X-rays
corresponds well to the first impulsive enhancement in the
optical lines, but we do not observe a significant continuum peak in
Figure \ref{fig:cont_evol} at this time. The maximum of the hard X-rays at 15:09:25 corresponds to a major peak in
both optical lines (15:09:30) and the continuum excess (15:09:30\,--\,15:09:51).  
The third emission line peak at 15:11:15 follows a 
significant excess continuum detection at 15:10:54 by one raster cycle (20 s) but does not 
readily have a corresponding peak in the hard X-rays.  Rather, the continuum peak at 15:10:54 may be associated 
with a cotemporal peak in the H$\gamma$ light curve at the same slit position but 
directly adjacent (0\arcsec.4\,--\,1\arcsec.2) to the umbral WL kernel along the slit in the direction of the plage 
ribbon.  This spatially adjacent flare enhancement (not shown in the figure) has two maxima in the H$\gamma$ light curve at 15:09:30\,--\,15:09:51 and 15:10:54 with comparable values to the maxima at the position of the umbral WL kernel (Figure \ref{fig:fermi1}a).  Interestingly, the second episode of continuum and line brightening at 15:10:54 is also not readily associated with a major, cotemporal
hard X-ray peak.  We note that a strong peak is also present in the H$\gamma$ line in the second slit position at 15:10:54 (see Figure \ref{fig:fermi1}b).

\begin{figure}[h]
\centering
\includegraphics[width = 5.0in]{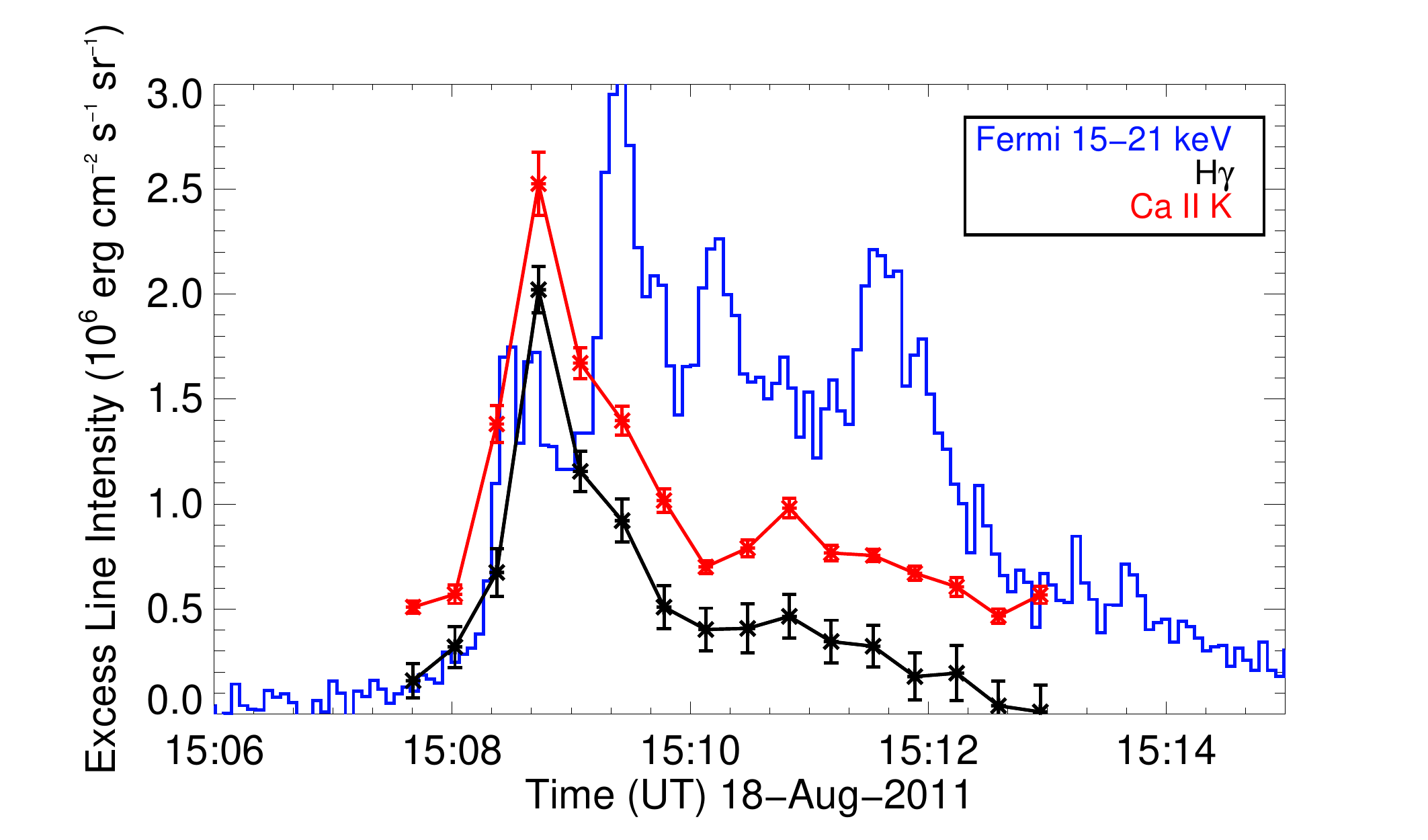}
\caption{The light curves of Ca \textsc{ii} K,
and H$\gamma$ for a region in the plage flare ribbon in the 18th slit position of the
raster compared to the Fermi $15-21$ keV light curve.  The emission line values were obtained from the 
ribbon intersected by the slit corresponding to the rightmost vertical white line in Figure \ref{fig:fov} (bottom left panel).  The average excess line intensity
is calculated over the three brightest spatial pixels in order to facilitate comparison to the umbral kernel (Figure \ref{fig:fermi1}).}
\label{fig:fermi2}
\end{figure}

The fourth major hard X-ray peak at 15:11:30 does not have a 
coincident peak in the optical emission lines originating from the umbral kernel 
and is probably associated with a different flaring area.  Searching the flaring region 
we find a possible association with an optical line increase in the 19th and 20th slit positions in the 
plage flare ribbon.  The first and second peaks of the Fermi light curve also correspond to 
peaks in the optical line emission originating from locations in the plage flare ribbon.  
An example light curve from a fixed spatial location (second white vertical line in Figure \ref{fig:fov})
in the plage flare ribbon is shown in Figure \ref{fig:fermi2}, which was obtained from the location of 
maximum optical line emission from the entire flare region.  
This light curve has a much more simple time evolution than the umbral kernel 
light curve (Figure \ref{fig:fermi1}a), but the line-integrated intensity is about twice as large even if the corresponding hard X ray burst is sensibly smaller than the following ones.
 However, the emission from the plage flare ribbon is not as spatially confined as the repeated optical line and H$\alpha+1.2$ \AA\ 
brightenings observed in the umbral kernel, which occur within a region confined to about 0\arcsec.5 (Figure \ref{fig:rasterHa}).  The single-peaked light curve morphology is  
consistent with the relatively rapid plage flare ribbon progression towards the weak field region in Figure \ref{fig:fov}.

For the umbral kernel, we compare the full-width at half-maximum (FWHM) of the light curves for H$\gamma$, Ca \textsc{ii} K, and hard 
X-rays in Figure \ref{fig:fermi1}a.  This measure gives a value known as $t_{1/2}$ which has been used for
characterizing the timescales of continuum and line emission for flares on dMe stars
\citep{Kowalski2013}.   Considering the entire light curve duration, the timescale of the hard X-rays is longer than the 
timescale of the optical lines because the additional major X-ray peak at 15:11:30 occurs without an optical line
counterpart at this spatial location.  We calculate the \emph{newly-formed} emission during the main peak at 15:09:30 by subtracting the flare emission at 15:08:50;  from this, we find that the $t_{1/2}$ values are 20~s, 60~s, and 60~s for the hard X-ray, H$\gamma$, and Ca \textsc{ii} K light curves, respectively.
Estimating the $t_{1/2}$ for the excess continuum in Figure \ref{fig:cont_evol} gives values ranging from 40-65~s, but this range is rather uncertain because 
the two significant continuum detections do not form a well-resolved light curve as for the emission lines.  
The significantly bright WL emission observed at 15:09:30 and 15:09:51 gives a lower limit of 20~s for the duration of the continuum excess, which is 
equal to the $t_{1/2}$ of the main hard X-ray peak.
This timing information will be important for guiding modeling efforts (Section \ref{sec:discussion}).

\subsection{The Balmer Decrement} \label{sec:decrement}

\begin{figure}[h]
\centering
\includegraphics[width = 5.0in]{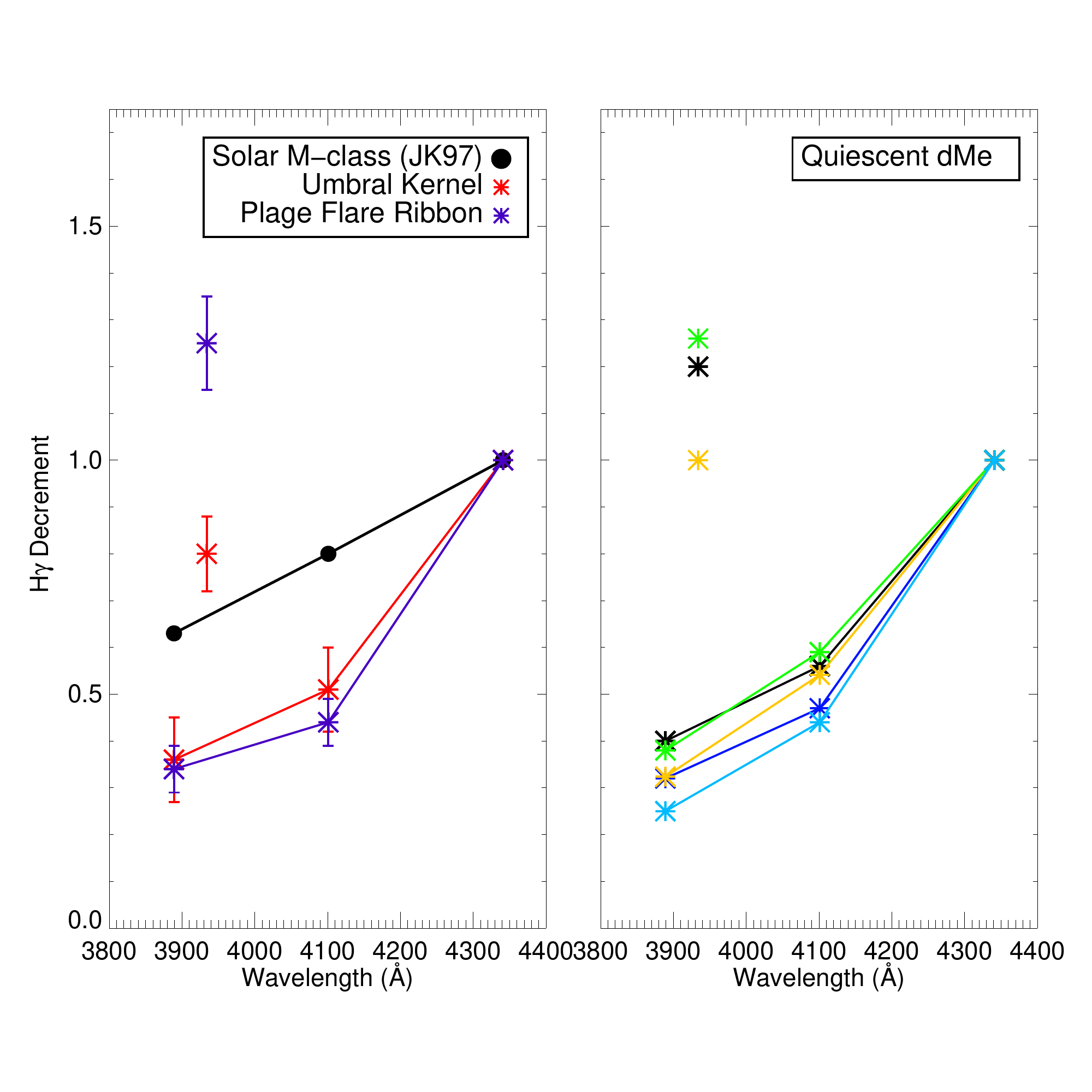}
\caption{\emph{Left:} The Balmer decrements for the plage flare ribbon at maximum line emission 15:08:44 (purple) and umbral
  kernel at 15:09:30 (red) compared to the decrements from the M7.7 solar flare presented in \cite{JK97}.  The decrements are shown as the wavelength-integrated emission in H8, Ca \textsc{ii} K,
  H$\delta$ divided by that in H$\gamma$; the lines connect only the hydrogen Balmer series.  The
  decrements were obtained from the emission line averaged over the
  same three pixels as used for continuum detection and analysis.
\emph{Right: } The Balmer decrements of quiescent dMe spectra obtained from the
  literature (black:  AD Leo from \cite{Hawley1991}; orange: AD Leo from
  \cite{Kowalski2013}; green:  UV Ceti from \cite{Phillips1988}; light blue: YZ CMi from \cite{Doyle1988}; dark blue:  AT Mic from \cite{Garcia2002}). }
\label{fig:decrement}
\end{figure}

A broad wavelength coverage, intensity-calibrated spectrum allows the relative intensity to be measured in each emission
line, giving the Balmer decrement, which is the ratio of the intensity of a particular Balmer line to that of another line (usually H$\gamma$) in the series. Due to the paucity of data needed to determine this parameter, the Balmer decrement is used infrequently in solar flare studies, but it is a powerful constraint on temperature, electron density, and H$\alpha$ optical depth \citep{Drake1980}, and can be used to test future radiative-hydrodynamic modeling of the flaring atmosphere. The results for the H$\gamma$ Balmer decrement are shown for the solar flare in Figure \ref{fig:decrement}, and compared with values observed in other solar and stellar environments.  

We display decrements for the umbral kernel at 15:09:30 (the maximum in Figure \ref{fig:fermi1}a) and for the plage flare ribbon at 18th slit position at 15:08:44 (the maximum in Figure \ref{fig:fermi2}).  We find that the H8/H$\gamma$ and H$\delta$/H$\gamma$ decrements are comparable in the umbral kernel and in the plage flare ribbon, but the Ca \textsc{ii} K/H$\gamma$ decrement exhibits a significant variation between the two regions. 
In the umbral kernel, the Ca \textsc{ii} K/H$\gamma$ decrement is less than 1 at 15:09:30 whereas in the plage flare ribbon, the decrement is greater than 1.  
At the maximum at 15:11:15 in the umbral kernel (not shown), the Ca \textsc{ii} K/H$\gamma$ decrement is 0.6($\pm0.07$).
The decrements from our C1.1 flare are compared to the early stage decrements from the M7.7 flare studied in \cite{JK97} and are found 
to be steeper.  

Coincidentally, decrements from our C1 solar flare are very close to the
decrements of a chromospherically active M dwarf (dMe) spectrum during quiescent times
without any moderate or major flares.  The decrements from various observed quiescent dMe spectra 
are shown in the right panel of Figure \ref{fig:decrement} (right panel).  These
decrements have been obtained from the literature including the dM3e star AD Leo from
\cite{Kowalski2013} and other measurements of dMe stars reported in
\cite{Hawley1991}.  The similarity between our C1.1 flare and the dMe
stars is unambiguous for the 
of H8/H$\gamma$ and H$\delta$/H$\gamma$ decrements, whereas the Ca
\textsc{ii} K/H$\gamma$ decrements fall between the C1.1 solar flare plage ribbon and umbral
kernel values.  Note, the spectra of the dMe stars represent the irradiance
 from the entire visible hemisphere of the star, whereas the solar
flare measurements represent only the emergent intensity at
$\mu=0.74$.

\subsection{Broadening of the Balmer Lines} \label{sec:broadening}
\begin{figure}[h]
\centering
\includegraphics[width = 3.0in]{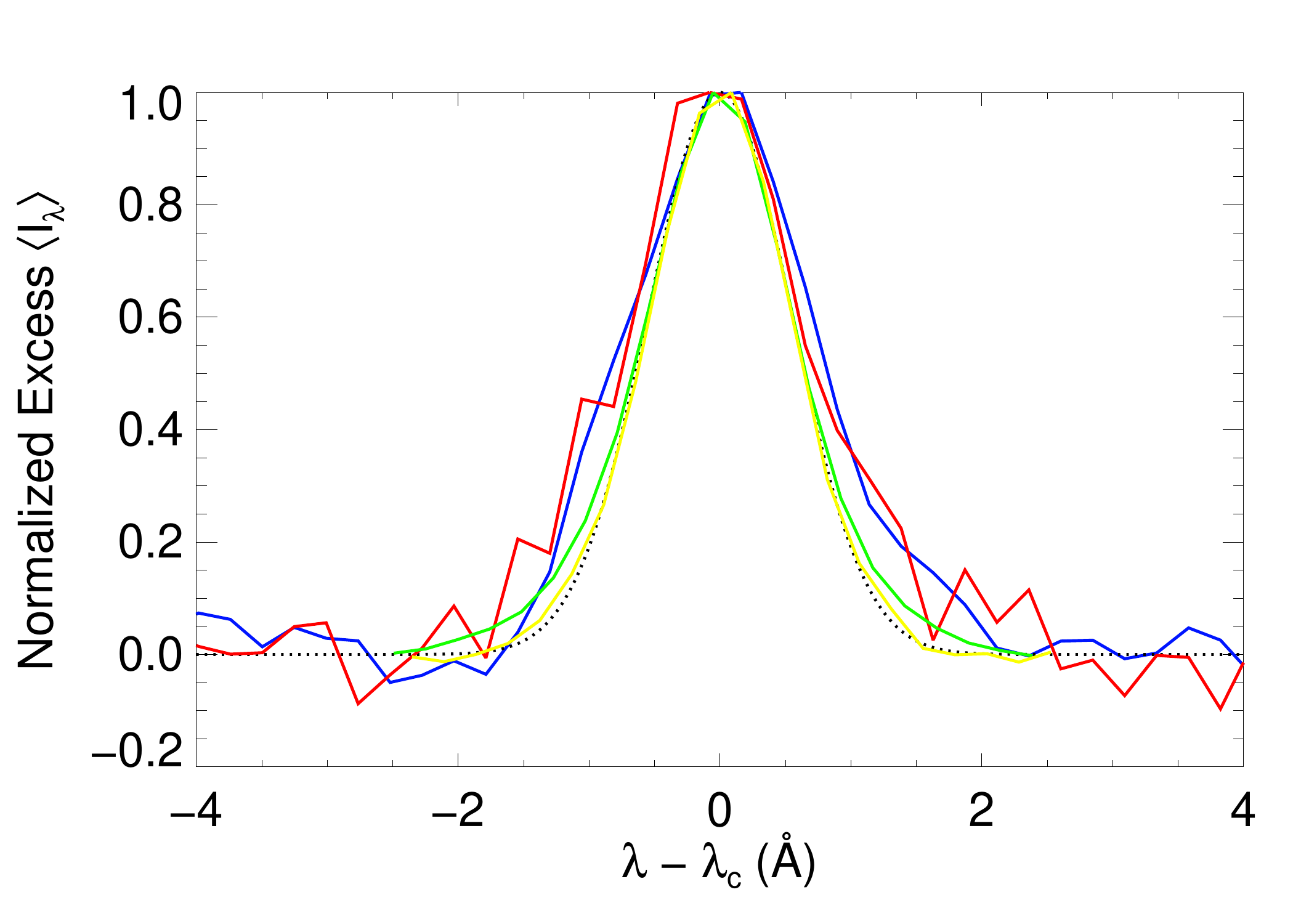}
\caption{The H8 line profile of the excess emission in umbral
  kernel (at 15:09:30; red) and plage flare ribbon (at 15:08:44; blue), 
normalized to their peak excess intensity.
  A gaussian with the instrumental FWHM of 1.3 \AA\ is shown as a dotted line, and 
the Ca \textsc{ii} K line profiles are shown at the same times and locations (plage flare ribbon as green, umbral kernel as yellow) as for the H8 profiles.  }
\label{fig:broadening}
\end{figure}

Symmetric broadening of hydrogen Balmer lines is thought to be at least partially due to the linear Stark 
effect from the ambient charge density in the flare chromosphere \citep{Svestka, Worden1984, JK97, Hawley1991}.  
Stark broadening theory predicts larger energy shifts in the highest energy levels of hydrogen, so we
examine the profiles of the highest order line with significant emission, H8 at $\lambda=3889$ \AA.  
We study the broadening at the same spatial locations and times as in Section \ref{sec:decrement} (at the light curve peaks of the umbral kernel and plage flare ribbion in Figures \ref{fig:fermi1}a, \ref{fig:fermi2}, respectively).
The line profiles normalized to their peaks
are shown in Figure \ref{fig:broadening}.
We find the FWHM of H8 is between $1.5-1.7$ \AA. 
The spectral resolution near H8 is at worst 1.3 \AA, which allows us to estimate an intrinsic FWHM of $\sim$1 \AA\ 
($\sigma_{\mathrm{observed}}^2 = \sigma_{\mathrm{instr}}^2 + \sigma_{\mathrm{intrinsic}}^2$ for a convolution of two gaussians). 
At the peak of the M7.7 flare reported in \cite{JK97}, the FWHM of
H8 was found to be 0.62 \AA\ or $\sim50$ km s$^{-1}$, which is less than the velocity 
width of $\sim$80 km s$^{-1}$ in our C1 flare.
 We also show the profiles of Ca \textsc{ii} K at these same 
times and locations;  this line is not affected by the linear Stark effect and generally 
has a profile that is close to the instrumental resolution, although there may be some broadening in the far wings.
Comparing to the Ca \textsc{ii} K broadening suggests that the H8 broadening is real, and 
a comparison of these two lines at 15:10:54 UT (not shown) in the umbral kernel does not reveal a significant difference.  
A meaningful understanding of the broadening mechanisms in our spectra
requires accurate modeling of the Stark profiles in addition to the contributions from thermal and turbulent
broadening, convolved with the instrumental profile.  As discussed
extensively by \cite{JK97}, the various ways that Stark broadening is
implemented in model codes can give somewhat ambiguous results for flare conditions; 
we will address this issue in a forthcoming modeling paper.  

\section{Summary and Discussion} \label{sec:discussion}

\subsection{Overview} \label{sec:overview_discussion}
We have reported an optical continuum detection and the emission line characteristics 
during a small C1.1 flare, observed during a test run of a customized
instrumental setup of the Horizontal Spectrograph on the Dunn Solar
Telescope.  The high spatial and temporal resolution of our observations allow 
us to clearly identify various portions of the flare that display different characteristics and evolution. 
In particular, within the flare ribbons, we detected a significant ($> 4\sigma$) excess in
optical continuum emission only in a tiny umbral kernel, with a diameter of about 0\arcsec.5 (350 km, Figures \ref{fig:raster2}-\ref{fig:rasterHa}).  
This umbral kernel exhibited repeated brightenings in the continuum and emission lines,
with an evolution that is generally similar to the early impulsive phase 
$15-21$ keV hard X-rays detected by Fermi but also with some noteable differences, including 
one major burst in the continuum excess (15:10:54) and optical lines (15:11:15) without cotemporal hard X-ray peaks
(Figures \ref{fig:fermi_goes}, \ref{fig:raster2}, \ref{fig:fermi1}).

No evidence of a significant continuum enhancement is found in any part of the plage flare ribbon, which develops concomitantly to the umbral one. However, several plage flaring kernels show enhanced chromospheric line emission in close temporal correlation with hard X ray peaks, with a rather impulsive character (e.g. Figure \ref{fig:fermi2}), which reflects the rapid spread of the plage ribbon within the weak magnetic field region.

\subsection{Balmer decrement and chromospheric line broadening} \label{sec:decrement_discussion}

Taking advantage of our broad spectral coverage, we calculated the Balmer decrements and line broadening both in the umbral kernel and at the location in the plage flare ribbon 
with the strongest line emission (Figures \ref{fig:decrement}, \ref{fig:broadening}). We found the decrements to be steeper and the broadening to be greater 
than during a M7 flare reported in the literature \citep{JK97}. These differences may be the result of both better spatial resolution of our data and of the different flare phase considered, with our data reflecting the very early impulsive phase vs. the more gradual one of \cite{JK97}.

We find an intriguing similarity of the  Balmer decrements in our flare to those of other stellar environments, and
speculate that there may be a similarity between the flaring conditions of a C1 solar flare and the lower atmospheric ``quiescent'' state of active M dwarf stars, which are known to show persistent hydrogen Balmer series and Ca \textsc{ii} K line emission. 
Quiescent coronal soft X-ray emission from active stars has been shown to be consistent with a superposition of many individual
flare events \citep{Gudel1997, Kashyap2002, Gudel2003}, and nonthermal turbulent broadening of quiescent transition region
lines has been interpreted as evidence of transition region explosive events or microflaring events occuring in regions of magnetic flux emergence \citep{Linsky1994}.  Persistent radio-emitting structures on dMe stars is indicative of the presence of nonthermal particles outside of major flaring events \citep{Osten2006}, and theoretical work indicates that particle acceleration and atmospheric heating is viable on active stars through a variety of mechanisms \citep{AH1998}.  Does the Balmer decrement 
 suggest that a fraction of the quiescent dMe chromospheric emission level can be attributed to the superposition 
of events similar to long duration C1 solar flares occurring in several active regions simultaneously, and continuously, on the stars?  
We leave this question open for a future investigation.

The Hydrogen line intensities provided by the slab model of \cite{Drake1980} can be used to reproduce the Balmer decrement curve shown in Figure \ref{fig:decrement}, and give a first indication of electron densities within the flaring region. By assuming a $T_e \sim 20,000$ K, we find that an H$\alpha$ optical depth of $\tau \sim 100-400$ and a density $n_e \sim 10^{11.5}-10^{12}$ cm$^{-3}$ are consistent with the measured  H$\delta$/H$\gamma$ and H8/H$\gamma$ decrements. Neither a higher electron density, or a lower H$\alpha$ optical depth can reproduce both decrements at once.

As mentioned in Section \ref{sec:broadening}, we plan to use these findings in a future work to estimate the effect of Stark broadening on the high order Hydrogen lines (H8 in particular), in comparison to the measurement of both Hydrogen and Ca \textsc{ii} K profiles widths.

\subsection{Continuum excess: intensity and spectral distribution} \label{sec:cont_excess_discussion}

The value and spectral distribution of the continuum excess measured in the flaring umbral kernel can provide constraints on the heating mechanisms acting on the lower atmosphere.
For example, a Balmer jump has been observed in several early flare spectra
\citep{Hiei1982, Neidig1983, Donati1985, Neidig1994}, which has led to the conclusion that 
the white-light continuum is comprised almost entirely of the Hydrogen
recombination spectrum (as opposed to an enhanced photospheric continuum) with an origin in 
the upper chromosphere \citep{Fletcher2007, Hudson1972}.  This has found further support in very recent observations obtained with the Interface Region Imaging Spectrograph, that highlighted the presence of near-ultraviolet (2813 \AA) continuum enhancement in some flaring kernels, fully consistent with hydrogen recombination Balmer continuum emission \citep{Heinzel2014}.

As reported in Section \ref{sec:reduction}, the uncertainties introduced by
chromatic aberration in our observations prevented an unambiguous determination of the detailed characteristics of the spectral slope at wavelengths in the Balmer continuum range ($\lambda < 3646$ \AA). 
However, we do note the apparent lack of a jump in
excess intensity or flare contrast at $\lambda < 3646$ \AA\ relative to the intensity at redder
wavelengths (Figure \ref{fig:balmerjump}).  A dominant optically thin 10,000~K spectrum would have produced a large Balmer jump ratio (i.e., the ratio of intensity at blue wavelengths to that at red wavelengths of 3646 \AA) that is $\sim 14$ \citep{Kunkel1970, Neidig1993}; we think that this should have been noticeable even with the 
data quality degraded due to chromatic aberration at $\lambda<3646$ \AA.  
It is possible that in our spectra the blending of Stark-broadened high-order Balmer lines\footnote{The spectral region from 3654-3674 \AA\ contains the rapidly converging hydrogen Balmer lines H23 through at least H40;  the center wavelengths of H23 and H24 are separated by $\sim$2.5 \AA\ whereas H39 and H40 are separated by only 0.5 \AA.} combined with the Stark broadening of the Balmer recombination edge could smear the Balmer jump making it non-detectable. 
Still, we note that other white light flares reported in the literature either did not display a Balmer jump, or had other properties not readily explained by a Hydrogen recombination spectrum. In particular,  a strong 
blue continuum emission at $\lambda < 4000$ \AA\ has been often reported \citep{Hiei1982,Neidig1984}, and shown by \cite{Donati1985} as the result of the blending of Stark-broadened high-order Balmer lines in a dense chromosphere. In the model of \cite{Donati1985}, this ``bump'' in the blue continuum peaks at $\lambda\sim3675$ \AA\, and becomes more prominent and shifts to redder wavelengths as electron density in the flare region increases.  Existing models of Stark broadening imply that such blue "continuum" emission originates from a location with electron density in excess of $10^{13}$~${\rm cm^{-3}}$ and electron temperature between 7,000 and 10,000~K. Interestingly, in our excess spectra we observe a relatively featureless, broad bump  peaking at $\lambda\sim$3675 \AA\ (Figure \ref{fig:balmerjump}).
Taken at face value, the electron densities inferred from interpreting this feature via Stark broadening appear at odd with those derived in Sect. \ref{sec:decrement_discussion}, unless more optically thin features such as the higher order Balmer lines probe different portions of the flaring atmosphere.  However, additional spectra that are not affected by chromatic aberration will be needed to confirm and understand this feature.

Following \cite{KerrFletcher}, we also compare our spectral 
data to a blackbody spectrum representing a photospheric flare continuum.
This, however, requires a well-resolved measurement of the intensity.  Although the umbral kernel is unresolved in the spectra,  the resolved area from IBIS H$\alpha+1.2$ \AA\ gives an actual spatial extent (0\arcsec.5) of the kernel that is not far below 
the resolution of the spectra (0\arcsec.67x0.8\arcsec);  therefore, we can give a lower limit on the radiation brightness temperature assuming a 
blackbody intensity.  The maximum excess$+$pre-flare intensity from $\lambda=4421-4451$ \AA\ in the umbral kernel at 15:09:30 gives a brightness temperature of $T_{\mathrm{rad}}\sim$5400 K for the flare, compared to $T_{\mathrm{rad}}\sim$5200 K for the pre-flare umbra.  This flare radiation temperature is similar to the optical color and brightness temperatures found in \cite{KerrFletcher} for an X-class flare and by \cite{Watanabe2013}.  
A photospheric temperature increase of only 200 K (also similar to that found in \cite{KerrFletcher}) likely implies a flare continuum emissivity dominated by H$^-$ emission processes (recombination and bremsstrahlung).  We note that this increase is far below that implied by a blackbody color temperature of $\sim$9000 K, recently found  in the Sun-as-a-Star, superposed epoch analysis of C class flares from \cite{Kretzschmar2011}.  
 The ratio of intensity at $\lambda=3914-3922$ \AA\ to the intensity 
at $\lambda=4421-4451$ \AA\ gives a color temperature (Section \ref{sec:WLresults}) in the umbral kernel for our data, but 
we do not expect the flare spectrum between these wavelengths to exhibit a Planckian shape for a small 
temperature increase of 200 K implied by the brightness temperature, due to similar complicated opacity effects that produce the pre-flare umbral spectrum.  There are relics of the background spectrum in the excess spectrum at the Ca
\textsc{ii} H and K absorption wings and at the G-band at $\sim$4300 \AA\ in Figure \ref{fig:spectrum1}.  These features may help constrain the origin of the emission using models that include wavelength-dependent opacities.

We finally turn to the flare contrast at different wavelengths, as defined in Section \ref{sec:WL_contrast}. This quantity can be largely affected by the spatial resolution, as discussed 
by \cite{Jess2008} for a C2.0 flare.  The flare contrast 
in Figure \ref{fig:spectrum1} was found to be $\sim10$\%, which is the average excess 
intensity relative to a nearby non-flaring (granulation) region away from the spot and between plage regions.  
Compared to some larger flares with spectra in the literature \citep[Figure 3 of ][]{Neidig1983}, the optical contrast values near $\lambda\sim3920$\AA\ are quite similar, but the contrast at the bluest wavelengths is significantly smaller.   It should be noted,
however, that these older spectra typically did not sample the brightest kernels. 
If instead we calculate the flare contrast relative to the pre-flare umbral intensity (Section \ref{sec:WLresults}, $I_o=I_{\rm{umbra}}$), we obtain values of $\sim20$\% or more for our flare.  However, the true spatial extent of the umbral kernel is only $A_{\mathrm{IBIS}}\sim$10$^{15}$ cm$^2$ (Section \ref{sec:WL_enhancement}) compared to the unresolved area from the spectra:  $A_{\mathrm{spec}} = 0.\arcsec67\mathrm{x}0.\arcsec8 \sim 3\times$10$^{15}$ cm$^2$.  This allows us to provide an estimate of the actual values of the flare contrast to be $\gtrsim30$\% ($\gtrsim$60\% relative to the pre-flare umbral intensity).
This adjusted value of the flare contrast (at $\lambda=3914-3922$\AA) is similar to the flare contrast value relative to the nearby granulation derived from a spatially resolved observation at $\lambda \sim 3954$ \AA\ for the C2.0 flare in \cite{Jess2008}.  The adjusted value of the flare contrast relative to pre-flare background is even consistent with the values obtained in several bright kernels at $\lambda=4275$\AA\ during the much larger X13 flare of 24-Apr-1984 \citep{Neidig1994}.
In making these adjustments we have multiplied by a factor of $A_{\mathrm{spec}}/A_{\mathrm{IBIS}} \sim 3$, which is consistent 
with summing the excess spectral intensity over the three spatial pixels (instead of averaging to produce $\left< I_{\lambda} \right>$).  
Applying the areal adjustment to the calculation of brightness temperature from $\lambda=4421-4451$\AA\ gives an increase of only 500 K (compared to an increase of 200 K without the adjustment), to a value of $T_{\mathrm{rad}}\sim$5700 K.  

\subsection{RHD modeling} \label{sec:RHD_discussion}
Detailed radiative-hydrodynamic (RHD) models of flares have been employed in the last years for a more rigorous
interpretation of  flare spectra. One such example is the RADYN code \citep[e.g.,][]{Carlsson1994, Carlsson1995, Carlsson1997}, which has been modified to incorporate flare energy deposition 
\citep{Hawley1994, Abbett1999, Allred2005}. The atmospheric dynamics \citep{Fisher1989} and optical continuum properties \citep{Cheng2010} depend strongly on the  
preflare atmospheric state (e.g., umbral vs. granulation), viewing angle ($\mu$), and 
parameters of the nonthermal electron spectrum (low-energy cutoff, $E_c$, power-law index, $\delta$, and energy flux) which is usually assumed to power the chromospheric emission.

Although the existing RHD models \citep{Abbett1999, Allred2005, Cheng2010} present results for generalized combinations
of heating parameters and preflare atmospheric states, they can provide some important insight  also for the case analyzed in this paper. 
 The models of \cite{Cheng2010} consider the flare contrast at several optical and infrared wavelengths
over a large parameter space of the nonthermal electron spectrum (used as the heating mechanism)
 while employing a model umbra for the 
preflare atmosphere, which would be most appropriate 
for our flare.  The model with $E_c=20$ keV, $\delta=5$, and nonthermal electron energy flux of 10$^{10}$ erg cm$^{-2}$ s$^{-1}$ (F10), 
at $\mu=0.95$, produces a contrast
of only 3\% at 4300\AA\ after 18~s of constant heating, which is far below the observed contrast of $20$\% (and the 
inferred corrected contrast of $\gtrsim 60$\%) in our flare.  A larger 
beam flux (10$^{11}$ erg cm$^{-2}$ s$^{-1}$, F11), higher low-energy cutoff ($40$ keV), and flatter spectral index ($\delta=3$) produce an acceptable value of the contrast at our viewing angle with a temperature increase of 300 K in the upper photosphere from chromospheric radiative-backwarming.  The models of 
\cite{Allred2005} produce an optical contrast (at $\lambda=5000$\AA) of 30\% for an F11 simulation relative to the granulation intensity, which is consistent with our inferred value (at $\lambda=4450$\AA) relative to granulation.  However, a large beam flux and a moderate to high ($20-40$ keV) low-energy cutoff would have produced also a large amount of Balmer continuum emission, and resulted in contrast values of $\sim50-230$\% relative to granulation at wavelengths just blueward of the Balmer edge \citep{Abbett1999, Allred2005}.  The lack of a strong Balmer continuum component (Section \ref{sec:WLresults} and Figure \ref{fig:balmerjump}) in our spectrum make it unlikely that such powerful level of energy deposition in the upper chromosphere and subsequent radiative-backwarming of the upper photosphere can explain the observations.

\subsection{Future work} \label{sec:future}

The preliminary analysis performed in the previous sections suggests that the observed properties of our flare are difficult to reconcile with simpler, static models, or with existing grids of RHD models. We thus plan to undertake detailed radiative-hydrodynamic models of this particular flare, utilizing our comprehensive set of observables to constrain the simulation.

The Fermi hard X ray data will be utilized to derive an
energy spectrum of nonthermal electrons for input to the models.  
From our optical spectral data we can also provide estimates for the time profile of heating, and 
a lower limit of the heating flux necessary to sustain the excess optical
flare emission.  The observed timescale ($t_{1/2}\sim$20~s and duration of 40~s) of the main hard X-ray burst can be used to  
guide the duration of the energy deposition time-profile for modeling the first significant continuum enhancement.   Using a simplifying, crude assumption that $\left<I_{\lambda,\mu}\right>$ is isotropic\footnote{In the optically thin, plane-parallel approximation, $F_{opt} = 2\pi \times 0.74\int \left<I_{\lambda,\mu=0.74,\mathrm{excess}}\right> d\lambda$},
 $F_{opt} = \pi\int
\left<I_{\lambda,\mu=0.74,\mathrm{excess}}\right> d\lambda$, integrated over the wavelengths of our observations, we obtain at 
the maximum of the umbral kernel (at 15:09:30), $F_{opt} \sim 5\times10^8$ erg cm$^{-2}$ s$^{-1}$.  
Adjusting this value by $A_{\mathrm{spec}}/A_{\mathrm{IBIS}}$ suggests the radiative flux to be at least $1.5 \times 10^9$ erg cm$^{-2}$ s$^{-1}$, which is still a lower limit because our spectrum has a limited wavelength coverage and does not include radiation emitted in the UV, red optical, and infrared.  Given the energy constraints from our data, future RHD models that aim to reproduce the optical emission during the main X-ray peak should carefully explore a range of nonthermal electron energy fluxes around the value of $\sim 10^9$ erg
cm$^{-2}$ s$^{-1}$, which is well feasible with the current generation of simulations \citep{Abbett1999, Cheng2010}. 
The models of \cite{Abbett1999} and \cite{Cheng2010} do not produce a significant optical contrast with such a low beam flux, but the particular combination of modeling parameters in these studies may not be appropriate for our flare.

Alternative heating mechanisms may be needed to explain the second continuum enhancement at 15:11 without an obvious, cotemporal X-ray peak.
If there is a relationship to the main X-ray peak, a delay of 90~s appears slightly too long for the lifetime of a downward-directed heated compression wave, or an Alfv\'enic disturbance,
to reach the conditions of optical line and continuum formation \citep{Fisher1989,Russell2013}.
Models of stochastic acceleration in magnetized turbulence predict that the relative amount of proton 
to electron acceleration increases in environments with a denser plasma, longer magnetic loops, or a weaker magnetic
field \citep{Petrosian2004, Emslie2004}, all of which may 
pertain to the atmospheric conditions during episodes of magnetic reconnection in the late impulsive phase.

\section{Conclusions} \label{sec:conclusions}
We observed a small-amplitude, long duration GOES C1.1 flare, and the observed and inferred 
optical properties (contrast and brightness temperature) appear similar to some X-class flares.  
Multiplying the area of the white-light kernel in our C1.1 flare by $F_{\mathrm{opt}}$ gives a power of $L_{\mathrm{opt}}=1.5\times10^{24}$ erg s$^{-1}$, 
which is comparable to the $1-8$\AA\ soft X-ray
luminosity of the entire flare region (Figure \ref{fig:fermi_goes}).  
Indeed, the soft X-ray emission in solar flares is only a minor fraction (1-10\%) of the total radiated 
energy \citep{Kretzschmar2011, Emslie2012},
and can vary largely from event to event \citep{NeidigKane1993}.  
What aspect of the flare energy release can explain such a variation in the soft X-ray response, and also in the 
apparent amount of Balmer continuum emission, while producing similar properties at optical wavelengths?

The broad spectral coverage of our data, in particular the rarely-observed blue wavelength range around the Balmer edge, provides an opportunity to confront the evolution of our flare with results from modern RHD models.  Very few events have been observed so comprehensively, providing a rigorous way to guide the models and assess their assumptions and results.   For example, the hard X ray clearly informs what kind of beam we can use, and for how long energy deposition is sustained.  The resolved area of the WL kernel constrains the heating flux, which is important for 
determining the flare dynamics and the contrast at wavelengths where Balmer continuum emission is expected.   Furthermore, the WL kernel is observed from the very beginning of the flare, so we can reliably compare the results of the models with spectra at the appropriate time.  Most observations of WL flares in the past were never observed in the very impulsive phase, and rarely did spectral observations sample the brightest kernels.  We can investigate the different properties of umbral and plage kernels that develop concomitantly to address why one develops WL and the other does not while producing much stronger chromospheric line emission.  The connection among particle acceleration (or other 
heating mechanisms) and the magnetic and atmospheric
 environment of the flare will likely be necessary to explore with detailed modeling, in order to explain the range of WL properties in the bluest wavelengths around the Balmer jump.

\acknowledgements
We want to thank the DST observers, D. Gilliam, M. Bradford, and J. Elrod, for their precious and patient assistance.  The authors thank Suzanne Hawley for useful discussions on the importance of obtaining solar spectra with broad wavelength coverage, K. Tolbert for assistance obtaining the Fermi data, M. Janvier for helpful suggestions and discussions, and an anonymous referee for helpful constructive comments.  The authors are pleased to acknowledge the support of the International Space Sciences Institute (ISSI) in Bern.  IBIS is a project of INAF/OAA with additional contributions from University of Florence and Rome and NSO. The National Solar Observatory is operated by the Association of Universities for Research in Astronomy, Inc., under a cooperative agreement with the National Science Foundation.   This research was supported by an appointment to the NASA Postdoctoral Program at the Goddard Space Flight Center, administered by Oak Ridge Associated Universities through a contract with NASA.  The research has received funding from the European Community's Seventh Framework Programme (FP7/2007-2013) under agreement \#606862 (F-CHROMA).  AFK acknowledges the National Solar Observatory for travel support.  LF acknowledges support from STFC grant ST/L000741/1.

\clearpage

\appendix 
\section*{Appendix A:  Intensity Calibration}\label{sec:appendix}
In this Appendix, we describe the detailed spectral reduction procedure and intensity calibration.  
De-focused quiescent solar spectra were obtained away from the active region
at UT 16:26 (airmass of 1.34) at the same heliocentric radius vector (0.68) and DST guider angle (119.8 deg) 
as the observations.  To isolate the CCD variations from the quiescent solar spectrum, we performed a median
filtering which resulted in a master flat field image.  This master
flat was divided out of all images.

Wavelength and intensity calibration was carried out using the disk-center absolute solar intensity spectrum 
obtained with the Fourier Transform Spectrometer (FTS) with spectral resolution R $=$ 350\,000 \citep{Neckel1999}.
The wavelength solution and spectral resolution were 
obtained by aligning the quiescent solar spectral features in the HSG spectra to the FTS
spectral features.  The HSG dispersions are 0.28 \AA\ pixel$^{-1}$, 0.24 \AA\
pixel$^{-1}$, 0.35 \AA\ pixel$^{-1}$, and 0.35 \AA\
pixel$^{-1}$ for the bluest to the reddest cameras, respectively.  We found that the
spectral resolution was approximately 0.9\,--\,1.2 \AA\ at 4300 \AA\ (R$\sim$4000) by convolving the
FTS spectrum with Gaussians of various widths.

To calibrate the active region spectra to an absolute intensity scale, we used the quiescent solar spectrum from UT 16:26 (with flat-field variations removed) as
the reference.  From this, we extracted an average solar spectrum over 10 spatial pixels, which was 
converted from counts spatial pixel$^{-1}$ wavelength pixel$^{-1}$ to counts
sr$^{-1}$ wavelength pixel$^{-1}$ by multiplying by $4.25 \times 10^{10}$
arcseconds$^2$ sr$^{-1}$ $\times$ 1 spatial pixel / 0\arcsec.39 $\times$ 1 / 0\arcsec.67 (pixel size along the slit, and slit size, respectively). 

The IRAF routines \emph{standard}
and \emph{sensfunc} were used to determine the instrumental
sensitivity.  These routines divide the reference solar spectrum
(in units of counts sr$^{-1}$ wavelength pixel$^{-1}$) by the exposure
time and user-defined wavelength bins in order to compare against the
FTS disk-center intensity spectrum averaged over the same wavelength bins.  The FTS
spectrum was multiplied by the limb darkening
($D$) corresponding to a radius vector of 0.68, and was converted to AB
magnitudes ($m_{\mathrm{AB}} = -2.5\mathrm{log}_{10} (I_{\nu}D) - 48.60$) for 
the IRAF routines.  The limb darkening was determined to be 0.83 
from equation 8 of \cite{Pierce1977} using $\mu$=0.74\footnote{We used
\cite{Sykes} second-order fit to  $\ln \mu$ given in \cite{Pierce1977};
the fifth order fit to ln $\mu$ in \cite{Pierce1977} gives a slightly
larger amount of limb darkening, 0.80. }.
The limb darkening is wavelength dependent, but over our spectral
range varies only by about 1\%, so we used
a constant limb darkening.  We used an atmospheric extinction curve
obtained from a nearby site at the Apache Point Observatory.  
The instrumental sensitivity was fit with a smooth
function in 10 \AA\ wide bins where the FTS spectrum was free of 
strong absorption lines.  

We used the resultant instrumental sensitivity function to calculate the wavelength-dependent conversion factor, $X(\lambda)$, which converts
the images in units of counts s$^{-1}$ sr$^{-1}$ \AA$^{-1}$ to intensity in units of ergs s$^{-1}$ cm$^{-2}$ sr$^{-1}$ \AA$^{-1}$
at the airmass (1.34) of the quiescent reference spectral observation.  The
conversion factor, $X(\lambda)$, which has units of ergs cm$^{-2}$ count$^{-1}$,
was adjusted for the wavelength-dependent atmospheric transmission at the airmass of the
target (active region) observation.  This was done by multiplying $X(\lambda)$ 
by $T(\lambda)_{ \rm{ref}}$ / $T(\lambda)_{\rm{targ}}$, where $T(\lambda)$ is the atmospheric transmission at a given airmass 
(i.e., $T(\lambda) = 10^{E(\lambda)\rm{sec~z}/ -2.5}$ where $E(\lambda)$ is the atmospheric extinction in units of magnitudes airmass$^{-1}$ and sec~z is the airmass of the observation). 
Before applying this final conversion
factor, the spectra were aligned to a common spatial orientation and
interpolated to a common pixel scale, 0\arcsec.39 pixel$^{-1}$.  Within each camera's spectral
range, wavelength-dependent shifts of 0.5 - 2 pixels were also applied to account for
differential refraction.  This calibration procedure was performed  
for all slit positions of all CCD's, resulting in a 2D image with 
wavelength and spatial pixel axes, having units of [$I_{\lambda, \mu = 0.74}$].

\begin{deluxetable}{cccccc}
\tabletypesize{\scriptsize}
\tablewidth{6.5in}
\tablecaption{Horizontal Spectrograph Instrumental Setup}
\tablehead{
\colhead{CCD} &
\colhead{Wavelength Range [\AA]} &
\colhead{Useable Wavelength Range$\dagger$ [\AA]} &
\colhead{Dispersion [\AA\ pix$^{-1}$]} &
\colhead{Exp Time [ms]} &
\colhead{Pixel Scale [\arcsec pix$^{-1}$]} 
}
\startdata
ccc06 & 3500-3790 & 3654-3674 & 0.28 & 500 & 0.39 \\
ccc01 & 3771-4020 & 3830-3978 & 0.24 & 40  & 0.34 \\
ccc07 & 3945-4306 & 4085-4125 & 0.35 & 20 & 0.48 \\
ccc08 & 4198-4559 & 4213-4553 & 0.35 & 10 & 0.48 \\
\enddata
\tablecomments{$\dagger$These correspond to wavelength ranges
that are useable for spectral characterization (e.g., slope determination).  For ccc06 and ccc07 we
display in the figures larger spectral ranges of 3600-3740 \AA\ and 4016-4200 \AA\ for only the purposes of white-light detection (see comments on chromatic aberration in Section \ref{sec:reduction}). }
\label{table:obsdetails}
\end{deluxetable}

\bibliographystyle{apj}
\bibliography{aug182011_WL_vfinal_rev_v2}

\end{document}